\documentclass[twocolumn,showpacs,preprintnumbers,amsmath,amssymb,superscriptaddress]{revtex4}

\usepackage{graphicx}
\usepackage{dcolumn}
\usepackage{bm}
\usepackage[figuresright]{rotating}
\usepackage{fancyhdr}
\usepackage{enumerate}
\usepackage{indentfirst}
\usepackage{xspace}
\usepackage{color}

\begin{document}

\title{\bf Measurement of single charged pion production in the charged-current interactions of neutrinos in a 1.3 GeV wide band beam}
\newcommand{\BCN}{\affiliation{Institut de Fisica d'Altes Energies, Universitat Autonoma de Barcelona, E-08193 Bellaterra (Barcelona), Spain}}
\newcommand{\BU}{\affiliation{Department of Physics, Boston University, Boston, Massachusetts 02215, USA}}
\newcommand{\UBC}{\affiliation{Department of Physics \& Astronomy, University of British Columbia, Vancouver, British Columbia V6T 1Z1, Canada}}
\newcommand{\UCI}{\affiliation{Department of Physics and Astronomy, University of California, Irvine, Irvine, California 92697-4575, USA}}
\newcommand{\SACLAY}{\affiliation{DAPNIA, CEA Saclay, 91191 Gif-sur-Yvette Cedex, France}}
\newcommand{\CNU}{\affiliation{Department of Physics, Chonnam National University, Kwangju 500-757, Korea}}
\newcommand{\DU}{\affiliation{Department of Physics, Dongshin University, Naju 520-714, Korea}}
\newcommand{\DUKE}{\affiliation{Department of Physics, Duke University, Durham, North Carolina 27708, USA}}
\newcommand{\GENEVA}{\affiliation{DPNC, Section de Physique, University of Geneva, CH1211, Geneva 4, Switzerland}}
\newcommand{\UH}{\affiliation{Department of Physics and Astronomy, University of Hawaii, Honolulu, Hawaii 96822, USA}}
\newcommand{\KEK}{\affiliation{High Energy Accelerator Research Organization(KEK), Tsukuba, Ibaraki 305-0801, Japan}}
\newcommand{\HIR}{\affiliation{Graduate School of Advanced Sciences of Matter, Hiroshima University, Higashi-Hiroshima, Hiroshima 739-8530, Japan}}
\newcommand{\INR}{\affiliation{Institute for Nuclear Research, Moscow 117312, Russia}}
\newcommand{\KOBE}{\affiliation{Kobe University, Kobe, Hyogo 657-8501, Japan}}
\newcommand{\KOR}{\affiliation{Department of Physics, Korea University, Seoul 136-701, Korea}}
\newcommand{\KYO}{\affiliation{Department of Physics, Kyoto University, Kyoto 606-8502, Japan}}
\newcommand{\LSU}{\affiliation{Department of Physics and Astronomy, Louisiana State University, Baton Rouge, Louisiana 70803-4001, USA}}
\newcommand{\MIT}{\affiliation{Department of Physics, Massachusetts Institute of Technology, Cambridge, Massachusetts 02139, USA}}
\newcommand{\MIYAGI}{\affiliation{Department of Physics, Miyagi University of Education, Sendai 980-0845, Japan}}
\newcommand{\NIIGATA}{\affiliation{Department of Physics, Niigata University, Niigata, Niigata 950-2181, Japan}}
\newcommand{\OKAYAMA}{\affiliation{Department of Physics, Okayama University, Okayama, Okayama 700-8530, Japan}}
\newcommand{\OSAKA}{\affiliation{Department of Physics, Osaka University, Toyonaka, Osaka 560-0043, Japan}}
\newcommand{\ROME}{\affiliation{University of Rome La Sapienza and INFN, I-000185 Rome, Italy}}
\newcommand{\SNU}{\affiliation{Department of Physics, Seoul National University, Seoul 151-747, Korea}}
\newcommand{\SOLTAN}{\affiliation{A.~Soltan Institute for Nuclear Studies, 00-681 Warsaw, Poland}}
\newcommand{\TOHOKU}{\affiliation{Research Center for Neutrino Science, Tohoku University, Sendai, Miyagi 980-8578, Japan}}
\newcommand{\SB}{\affiliation{Department of Physics and Astronomy, State University of New York, Stony Brook, New York 11794-3800, USA}}
\newcommand{\TUS}{\affiliation{Department of Physics, Tokyo University of Science, Noda, Chiba 278-0022, Japan}}
\newcommand{\KAM}{\affiliation{Kamioka Observatory, Institute for Cosmic Ray Research, University of Tokyo, Kamioka, Gifu 506-1205, Japan}}
\newcommand{\RCCN}{\affiliation{Research Center for Cosmic Neutrinos, Institute for Cosmic Ray Research, University of Tokyo, Kashiwa, Chiba 277-8582, Japan}}
\newcommand{\TRIUMF}{\affiliation{TRIUMF, Vancouver, British Columbia V6T 2A3, Canada}}
\newcommand{\VAL}{\affiliation{Instituto de F\'{i}sica Corpuscular, E-46071 Valencia, Spain}}
\newcommand{\UW}{\affiliation{Department of Physics, University of Washington, Seattle, Washington 98195-1560, USA}}
\newcommand{\WARSAW}{\affiliation{Institute of Experimental Physics, Warsaw University, 00-681 Warsaw, Poland}}

% to make affiliation numbers in order
\BCN
\BU
\UBC
\UCI
\SACLAY
\CNU
\DU
\DUKE
\GENEVA
\UH
\KEK
\HIR
\INR
\KOBE
\KOR
\KYO
\LSU
\MIT
\MIYAGI
\NIIGATA
\OKAYAMA
\OSAKA
\ROME
\SNU
\SOLTAN
\TOHOKU
\SB
\TUS
\KAM
\RCCN
\TRIUMF
\VAL
\UW
\WARSAW

\author{A.~Rodriguez}\BCN 
\author{L.~Whitehead}\SB 
\author{J.~L.~Alcaraz}\BCN                
%\author{E.~Aliu}\BCN                
\author{S.~Andringa}\BCN 
\author{S.~Aoki}\KOBE 
\author{J.~Argyriades}\SACLAY 
\author{K.~Asakura}\KOBE 
\author{R.~Ashie}\KAM 
\author{F.~Berghaus}\UBC
\author{H.~Berns}\UW 
% ...
\collaboration{The K2K Collaboration}\noaffiliation

\begin{abstract}
% Editors: Lisa

Single charged pion production in charged-current muon neutrino interactions with carbon is studied using data collected in the K2K long-baseline neutrino experiment.  The mean energy of the incident muon neutrinos is 1.3 GeV.  The data used in this analysis are mainly from a fully active scintillator detector, SciBar.  The cross section for single $\pi^{+}$ production in the resonance region ($W<2$~GeV/$c^2$) relative to the charged-current quasi-elastic cross section is found to be 0.734 $^{+0.140}_{-0.153}$.  The energy-dependent cross section ratio is also measured.  The results are consistent with a previous experiment and the prediction of our model.

\end{abstract}

\maketitle

\section{Introduction}
\label{sec:introduction}
% Editors: Michel, Federico
Single charged pion production in the interactions of neutrinos with target material is dominated by a resonance process in the neutrino energy region of a few GeV.  In this process, production of a baryon resonance $N^{\ast}$ is followed by its prompt decay to a nucleon-pion final state. For charged-current (CC) interactions, the process can generically be written as $\nu_{\ell} N \rightarrow \ell N^{\ast},N^{\ast} \rightarrow N'\pi$, where $\ell$ is $e$, $\mu$, or $\tau$.  A similar reaction holds for the corresponding neutral-current (NC) process. Overall, there are six (eight) channels allowed in charged-current (neutral-current) interactions of neutrinos and antineutrinos.

The experimental investigation of resonant single charged pion production via CC interactions of neutrinos was first carried out in the 1960s and 1970s~\cite{Block:1964gj,Budagov:1969pw,Campbell:1973wg,Barish:1978pj,Bell:1978qu,Bell:1978rb,Lerche:1978cp} and was comprehensively described from a phenomenological point of view shortly after (Rein and Sehgal~\cite{Rein:1980wg}; Feynman, Kislinger and Ravndal~\cite{Feynman:1971wr}; Schreiner and von Hippel~\cite{Schreiner:1973mj}).
More experimental results for a variety of nuclear targets and neutrino energy regimes have been collected
in more recent years~\cite{Allen:1980ti,Radecky:1981fn,Kitagaki:1986ct,Allen:1985ti,Ammosov:1988xb,Grabosch:1988gw,Allasia:1990uy}. This paper presents the results obtained from the K2K experiment on single pion production with a carbon target in the region of hadronic invariant mass ($W$) typically associated with resonance production, $W<2$~GeV/$c^2$.  This result improves our knowledge of these reactions in the few-GeV neutrino energy regime, which is relevant for several future neutrino experiments. The experimental input from K2K and other neutrino experiments is of great importance to validate and tune models of neutrino excitation of baryon resonances, as well as models describing the effects due to the nuclear medium. This input is particularly important in the $\sim$1 GeV neutrino energy regime, where charged-current resonant single pion production
accounts for the second largest contribution to the total neutrino cross section after charged-current quasi-elastic (CCQE) scattering.

In addition, a better understanding of the single $\pi^+$ production mechanisms is of critical importance for answering fundamental questions that can be addressed with neutrino oscillation experiments. We give two examples. One fundamental question that is being probed with an increasing level of accuracy is if the ``atmospheric'' mixing angle $\theta_{23}$ is equal to $\pi/4$, thus providing maximal mixing in the ($\nu_{\mu},\nu_{\tau}$) sector. This question is
being addressed by looking for $\nu_{\mu}$ disappearance in long-baseline neutrino oscillation experiments with conventional neutrino beams. The main background to the muon neutrino induced CCQE signal is single $\pi^+$ production where the pion is not seen in the neutrino detector. Large uncertainties on this background limit the
accuracy with which this question can be answered. A second fundamental question that remains to be answered concerns the value of the small mixing angle $\theta_{13}$, parametrizing sub-leading neutrino oscillations and setting the scale for possible CP violation in the lepton sector. This question can be addressed, among other possibilities, via $\nu_{\mu}\to\nu_e$ and $\nu_e\to\nu_{\mu}$ searches at the atmospheric scale with superbeams and high-energy $\beta$-beams (neutrinos from boosted ion decays), respectively. Major backgrounds to these searches are resonant $\nu_{\mu}$ NC $\pi^0$ production and resonant $\nu_e$ neutral-current $\pi^+$ production, respectively~\cite{Bandyopadhyay:2007kx}. Both background processes are related, via isospin and electroweak relations, to the resonant $\nu_{\mu}$ CC $\pi^+$ production processes discussed in this paper.

The paper is organized as follows. Section \ref{sec:experiment} describes the neutrino beam and neutrino detectors. Section \ref{sec:simulation} describes the simulation of the experiment, focusing on the neutrino interaction simulation and the detector response. Section \ref{sec:analysis1} describes the cross section analysis, including the event selection and classification and the method used to extract the relative cross section. Section \ref{sec:systematics} describes the systematic uncertainties affecting the results of this paper, which are given in Section \ref{sec:results} with a comparison to the neutrino interaction simulation and existing results. Conclusions
are given in Section \ref{sec:conclusions}.

\section{Experimental Setup}
\label{sec:experiment}
% Editors: Michel, Federico

\subsection{Neutrino Beam}
\label{subsec:experiment_beam}
% Editors: Michel, Federico

%Beamline, integrated pot and run period, results of
%spectrum fit for neutrino flux, etc.
The KEK to Kamioka (K2K) experiment~\cite{Ahn:2006zz,Aliu:2004sq,Ahn:2002up,Ahn:2001cq} is a long-baseline neutrino oscillation experiment in which a beam of muon neutrinos is created at KEK in Tsukuba, Japan and sent to the Super-Kamiokande detector~\cite{Fukuda:2002uc}, located in Kamioka, Japan, 250 km away.  To produce the neutrino beam, protons are accelerated by the KEK proton synchrotron to a kinetic energy of 12 GeV. After acceleration, all protons are extracted in a single turn to the neutrino beam line. The duration of an extraction, or spill, is 1.1 $\mu$s, and each spill contains 9 bunches of protons with a 125 ns time interval between them. The cycle is repeated every 2.2 s. 
The direction and yield of the neutrino beam are checked by monitoring the muons produced by pion decay and occasionally monitoring the pions focused by the horn magnets. Over the duration of the K2K experiment, a total of $104.9\times10^{18}$ protons were delivered to the target to generate the neutrino beam. The SciBar detector, which will be described below, took data from October 2003 until November 2004; a total of 21.7$\times$10$^{18}$ protons on target for analysis were accumulated during this time period.

\begin{figure}[ht]
\begin{center}
  \includegraphics[width=0.45\textwidth]{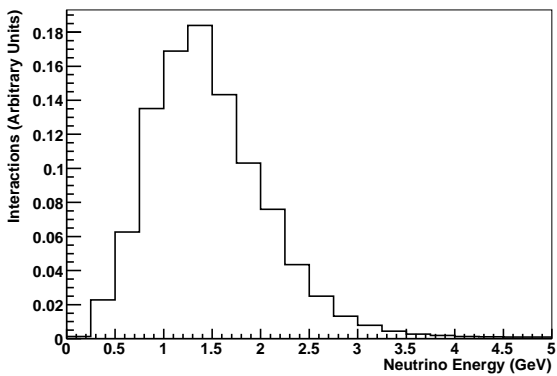}
\caption{The simulated neutrino energy spectrum for all muon neutrino interactions in the fiducial volume of the SciBar detector.}
\label{spectrum}
\end{center}
\end{figure}

We use a Monte Carlo (MC) simulation to predict the properties of our neutrino beam~\cite{Ahn:2006zz}.  According to this simulation, the beam at the near detector is about 97.3\% pure $\nu_{\mu}$ with a mean energy of 1.3 GeV.  The rest of the beam is mainly $\nu_{e}$ ($\nu_{e}/\nu_{\mu} \sim 0.013$) and $\overline{\nu}_{\mu}$ ($\overline{\nu}_{\mu}/\nu_{\mu} \sim 0.015$).  Data from all the near detectors are used to fine-tune the simulated neutrino energy spectrum.
Figure \ref{spectrum} shows the simulated energy spectrum for all muon neutrino interactions in the fiducial volume of the SciBar detector.

\subsection{Neutrino Detectors at KEK}
\label{subsec:experiment_neardet}
% Editors: Michel, Federico

%Near detector general description, especially MRD since we use
%it. SciBar next.

\begin{figure}[ht]
\begin{center}
  \includegraphics[width=0.45\textwidth]{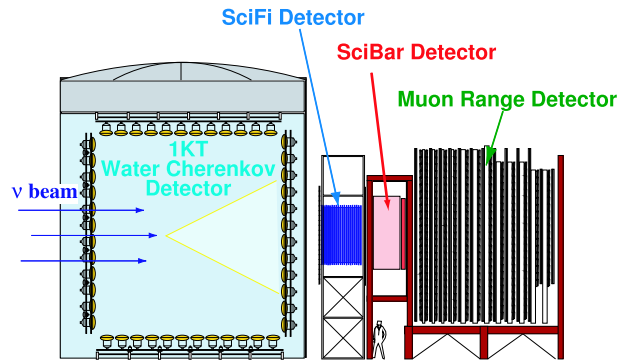}
\caption{(Color online) Schematic view of the near neutrino detector.}
\label{nd}
\end{center}
\end{figure}

The near neutrino detector system is located 300 m downstream from the proton target. The purpose of the near detector is to measure the direction, flux, and energy spectrum of neutrinos at KEK before oscillation; the near detector can also be used for measurements of neutrino-nucleus cross sections.  A schematic view of the near detector is shown Figure \ref{nd}. The near detector consists of a one kiloton water \v{C}erenkov detector (1KT)~\cite{Nakayama:2004dp}, a scintillating-fiber/water target tracker (SciFi)~\cite{Suzuki:2000nj}, a fully active scintillator-bar tracker (SciBar, since Oct. 2003), and a muon range detector (MRD). We describe in this section the SciBar and MRD detectors; data taken from both detectors is used in this analysis.

\subsubsection{SciBar Detector}
\label{subsec:experiment_scibar}
% Editors: Lisa

% Detector description, also xtalk and DAQ (eg, hit threshold)
% effects, and also reconstruction algorithms here.

\begin{figure}[htp]
  \begin{center}
    \includegraphics[width=0.45\textwidth]{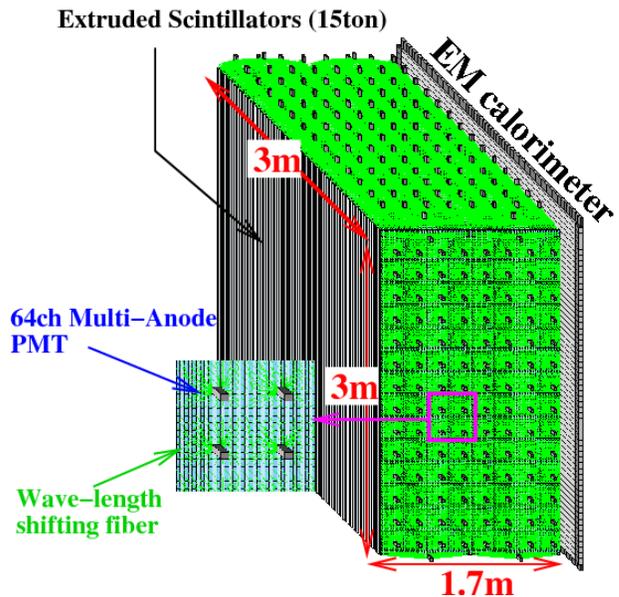}
  \end{center}
  \caption{\label{fig:scibar}(Color online) Diagram of SciBar.}
\end{figure}

SciBar~\cite{Nitta:2004nt,Yamamoto:2005cy} consists of 14,848 scintillating bars.  Groups of 116 bars are arranged horizontally or vertically to make one plane.  One layer consists of one horizontal plane and one vertical plane; there are 64 layers in total.  SciBar is a fully active detector.  The total volume is 1.7 m $\times$ 3 m $\times$ 3 m, for a total mass of $\sim$15 tons. Figure \ref{fig:scibar} shows a diagram of the SciBar detector.

The bars are made of polystyrene (C$_8$H$_8$), PPO (1\%), and POPOP (0.03\%).  Each bar is 1.3 cm $\times$ 2.5 cm $\times$ 300 cm and has a 0.25 mm thick reflective coating made of polystyrene containing 15\% of TiO$_2$ by weight.  The peak of the emission spectrum for the scintillator is at 420 nm.

A 1.5 mm diameter wavelength shifting (WLS) fiber (Kuraray Y11(200)MS) is inserted in a 1.8 mm hole in each bar to guide the scintillation light to multi-anode photomultiplier tubes (MAPMTs).  The average attenuation length of the WLS fibers is approximately 350 cm.  The fibers have a polystyrene core (refractive index $n$=1.56), an inner cladding of acrylic ($n$=1.49), and an outer cladding of polyfluor ($n$=1.42).  The absorption peak for the fibers is at 430 nm (matching the emission peak for the scintillator), and the emission peak for the fibers is at 476 nm.

The scintillation light is detected by Hamamatsu H8804 MAPMTs.  Each MAPMT has 64 channels arranged in an 8$\times$8 array.  Each pixel is 2 mm $\times$ 2 mm.  The cathode material is bialkali, with a quantum efficiency of 21\% at a wavelength of 390 nm.  The cathode is sensitive to wavelengths between 300 and 650 nm.  The basic properties such as gain and linearity are measured for each channel before installation.  The high voltage of each MAPMT was tuned so that the average gain of the 64 channels was 6$\times$10$^5$.  The non-linearity of the output signal vs. input charge is 5\% at 200 photoelectrons (p.e.) at a gain of 5$\times$10$^5$.  Crosstalk in the MAPMT is approximately 3\% in neighboring channels.  Groups of 64 fibers are bundled together and glued to a connector to be precisely aligned with the pixels of the MAPMT.

SciBar's readout system~\cite{Yoshida:2004mh} consists of a front-end electronics board (FEB) attached to each MAPMT and a back-end VME module.  The front-end electronics uses VA/TA ASICs.  The VA is a 32-channel pre-amplifier chip with a shaper and multiplexer.  The TA provides timing information by taking the ``OR'' of 32 channels.  Each FEB uses two VA/TA packages to read 64 analog signals and two timing signals for each MAPMT.  Each back-end VME board controls the readout of eight FEBs.  Flash ADCs are used to digitize the charge information, and TDCs are used to process the timing information.  The pedestal width is approximately 0.3 p.e, and the timing resolution is 1.3 ns.

In order to monitor and correct for gain drift during operation, SciBar is equipped with a gain calibration system using LEDs~\cite{Hasegawa-thesis}.  The system shows that the gain is stable within 5\% for the entire period of operation.  Cosmic ray data collected between beam spills are used to calibrate the light yield of each channel.  The average light yield per bar is approximately 20 p.e. for a minimum ionizing particle.  The light yield is stable within 1\% for the whole period of operation, after taking the gain variation into account.  Pedestal, LED, and cosmic ray data are taken simultaneously with beam data.

A crosstalk correction is applied to both data and MC before event reconstruction.  Let $M$ be the 64$\times$64 crosstalk matrix, where $M_{ij}$ is the fraction of channel $j$'s signal that migrates to channel $i$ due to crosstalk.  If $q_i$ is the charge in channel $i$ before crosstalk and $q'_{i}$ is the charge in channel $i$ after crosstalk, then
\begin{equation}
q'_{i} = \sum_{j}M_{ij}q_{j}.
\end{equation}
The crosstalk correction is just the inverse process,
\begin{equation}
q_{i} = \sum_{j}M^{-1}_{ij}q'_{j}.
\end{equation}
After the crosstalk correction, hit scintillator strips with at least two p.e. (corresponding to about 0.2 MeV) are selected for tracking.  Charged particles are reconstructed by looking for track projections in each of the two-dimensional (2D) views ($x$-$z$ and $y$-$z$) using a cellular automaton algorithm~\cite{Glazov:1993ur}. The $z$-axis is the axis perpendicular to the detector planes and is offset approximately 1 degree from the beam direction. Three-dimensional (3D) tracks are reconstructed by matching the $z$-edges and timing information of the 2D tracks.  Reconstructed tracks are required to have hits in at least three consecutive layers.  Therefore, the minimum length of a reconstructible track is 8 cm in the beam direction, which corresponds to a momentum threshold of 450 MeV/$c$ for protons.  The reconstruction efficiency for an isolated track at least 10 cm long in the beam direction is 99\%.   The efficiency is lower for multiple track events due to overlapping of tracks in one or both views.

Just downstream of SciBar is an electromagnetic calorimeter (EC).  The purpose of the EC is to measure the electron neutrino contamination in the beam and $\pi^{0}$ production in neutrino interactions.  The EC consists of one plane of 30 horizontal bars and one plane of 32 vertical bars.  The bars were originally made for the CHORUS neutrino experiment at CERN~\cite{Buontempo:1994yp}.  Each bar is a sandwich of lead and scintillating fibers.  The two planes are each 4 cm thick with cross sectional areas of 2.7 m $\times$ 2.6 m and 2.6 m $\times$ 2.5 m, respectively.  The EC adds an additional 11 radiation lengths (the main part of SciBar is about four radiation lengths).  The energy resolution for electrons is 14\%/$\sqrt{E\textnormal{(GeV)}}$ as measured by a test beam~\cite{Buontempo:1994yp}.

\subsubsection{Muon range detector (MRD)}
\label{subsubsec:MRD}
The MRD~\cite{Ishii:2001sj} is the most downstream detector. It consists of 12 layers of iron in between 13 layers of vertical and horizontal drift-tubes.  Each layer is approximately 7.6 m $\times$ 7.6 m.  To have good energy resolution for the whole energy spectrum, the four upstream iron layers are each 10 cm thick, while the other eight planes are 20 cm thick, for a total iron thickness of 2 m.  The total iron thickness ensures containment of forward-going muons depositing up to 2.8 GeV of energy in the MRD alone. There are 6632 aluminum drift tubes filled with P10 gas (Ar:CH$_4$ = 90\%10\%).  The total mass of the iron is 864 tons, and the mass of the drift tubes is 51 tons.

The MRD is used to identify muons produced in the upstream detectors.  The energy and angle of the muon can be measured by the combination of the MRD and the other fine-grained detectors.  

The MRD tracking efficiency is 66\%, 95\%, and 97.5\% for tracks that traverse one, two, and three iron layers, respectively; for longer tracks, the efficiency approaches 99\%.  The range of a track is estimated using the path length of the reconstructed track in iron.  The muon energy is then calculated by the range of the track.  The uncertainty in the muon energy due to  differences among various calculations of the relationship between muon energy and range is quoted to be 1.7\%.  The uncertainty in the weight of the iron is 1\%.  Thus, the systematic error in the MRD energy scale is conservatively quoted to be the linear sum of these uncertainties, 2.7\%, as in the K2K oscillation analysis~\cite{Ahn:2006zz}.

\section{Simulation}
\label{sec:simulation}
% Editors: Michel, Federico

\subsection{Neutrino Interactions}
\label{subsec:simulation_interactions}
% Editors: Lisa

% Description of NEUT, neutrino interaction reweightings
% (coherent, DIS) and final state interactions within
% the nucleus.

K2K uses the NEUT~\cite{Hayato:2002sd} neutrino interaction simulation library.  In NEUT, the following charged-current neutrino interactions are simulated: quasi-elastic scattering ($\nu_{\ell} N \rightarrow \ell N'$), single meson production ($\nu_{\ell} N \rightarrow \ell N' m$), coherent $\pi$ production ($\nu_{\ell} ^{16}{\rm O}(^{12}{\rm C},^{56}{\rm Fe}) \rightarrow \ell \pi ^{16}{\rm O}(^{12}{\rm C},^{56}{\rm Fe})$), and deep inelastic scattering ($\nu_{\ell} N \rightarrow \ell X$).  The corresponding neutral-current interactions are also simulated.  In these reactions, $N$ and $N'$ are nucleons, $\ell$ is a lepton, $m$ is a meson, and $X$ is a system of hadrons. For neutrino interactions occurring inside a nucleus, the interactions of the outgoing particles inside the nucleus are also considered. Table \ref{tab:nuint} shows the fraction of interactions in SciBar that are expected to be quasi-elastic, single pion, etc. according to the simulation.  Sections \ref{subsubsec:singlemeson} and \ref{subsubsec:otherint} will provide descriptions of neutrino interactions with free nucleons, while Section \ref{subsubsec:nuclear} deals with how the effects of the nuclear medium are taken into account for those interactions.
\begin{table}[htp]
\caption{Neutrino Interactions in SciBar}
\label{tab:nuint}
\begin{tabular}{l|c}
\hline
\hline
\textbf{~~Interaction type}&\textbf{Percent of Total}\\
\hline
\hline
\textbf{Charged-current (CC)}&\textbf{72\%}\\
~~~~~~CC quasi-elastic&~~~~32\%\\
~~~~~~CC single pion production&~~~~29\%\\
~~~~~~CC deep inelastic scattering&~~~~9\%\\
~~~~~~CC (other)&~~~~2\%\\
\textbf{Neutral-current (NC)}&\textbf{28\%}\\
\hline
\hline
\end{tabular}
\end{table}

\subsubsection{Single meson production}
\label{subsubsec:singlemeson}
Rein and Sehgal's model is used to simulate the production of single pions via baryon resonance excitation~\cite{Rein:1980wg,Rein:1987cb}.  In this model, the cross section for each $\ell N \pi$ final state is calculated as a coherent superposition of all the possible contributing resonances.

The differential cross section for the production of a single resonance with mass $M$ and width $\Gamma$ is given by
\begin{align}
&\frac{d^2\sigma}{dq^{2}dW} = \frac{1}{32\pi m_{N}E_{\nu}^{2}}\frac{1}{2} \times \notag\\
&\sum_{spins} |T(\nu N \rightarrow \ell N^{*})|^{2} \frac{1}{2\pi}\frac{\Gamma}{(W-M)^2 + \Gamma^{2}/4},
\end{align}
where $q^{2}$ is the square of the lepton momentum transfer, $W$ is the invariant mass of the produced baryon, $m_{N}$ is the nucleon mass, and $E_{\nu}$ is the incident neutrino energy.
Rein and Sehgal use the relativistic harmonic oscillator quark model of Feynman, Kislinger, and Ravndal~\cite{Feynman:1971wr} to calculate the transition matrix elements $T(\nu N \rightarrow l N^{*})$ from a ground state nucleon to a baryon resonance.  There is only one parameter in the model to be newly adjusted by neutrino scattering experiments, the axial vector mass, $M_{A}$.  It is set to 1.1 GeV/$c^2$ in our simulation.  The model for the decay of each resonance to an $N\pi$ final state uses experimental input for resonance mass, resonance width, and $N\pi$ branching ratios.

The cross section for the production of each $N\pi$ final state can be found by summing the contributions from each resonance, using appropriate factors determined by isospin Clebsch-Gordon rules.  The interference of overlapping resonances is taken into account.  Our simulation considers 18 baryon resonances below an invariant mass ($W$) of 2~GeV/$c^2$, as well as a non-resonant background contribution.

We use Rein and Sehgal's method for the pion angular distribution for the dominant resonance, $P_{33}$(1232). For the other resonances, the angular distribution of the pion is isotropic in the resonance rest frame.  The MC prediction for the $\pi^{+}$ angular distribution for the $\nu p \rightarrow \mu^- p \pi^+$ mode agrees well with a measurement~\cite{Kitagaki:1986ct}.

\begin{figure}[htp]
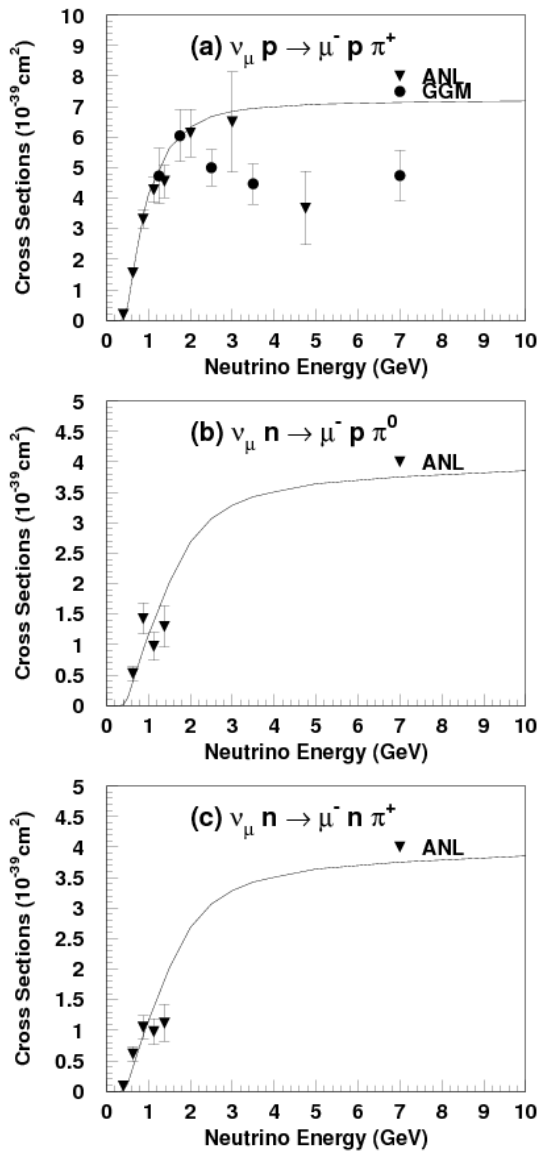

  \begin{center}
    \includegraphics[width=0.4\textwidth]{figures/single-pion1.eps2}
    \includegraphics[width=0.4\textwidth]{figures/single-pion2.eps2}
    \includegraphics[width=0.4\textwidth]{figures/single-pion3.eps2}
  \end{center}
  \caption{\label{fig:1pi_xsec} The cross sections for charged-current single pion production calculated in NEUT compared to experimental measurements, \mbox{({$\bullet$})GGM\protect~\cite{Lerche:1978cp}},\mbox{({$\blacktriangledown$})ANL\protect~\cite{Radecky:1981fn}}.}
\end{figure}

Figure \ref{fig:1pi_xsec} shows the calculated cross sections for the 3 modes of CC single pion production by muon neutrinos on nucleons compared with experimental measurements.

Our simulation also considers the production of single $K$ and $\eta$ using the same model.

\subsubsection{Other Neutrino Interactions}
\label{subsubsec:otherint}
Our simulation uses the formalism for CCQE scattering off a free nucleon described by Smith and Moniz~\cite{Smith:1972xh}. The axial-vector mass ($M_{A}$) for the CCQE interaction is set to 1.1 GeV/$c^2$. %based on near detector data~\cite{Ahn:2002up}.

For deep inelastic scattering (DIS), we use the GRV94 nucleon structure functions~\cite{Gluck:1994uf}, with a correction in the small $q^2$ region developed by Bodek and Yang~\cite{Bodek:2002vp}.  For the simulation of DIS interactions in which the hadronic invariant mass, $W$, is greater than 2~GeV/$c^2$, we use the PYTHIA/JETSET~\cite{Sjostrand:1993yb} package.  For the $W<2$ GeV/$c^2$ region, we use a custom library~\cite{Nakahata:1986zp} based on experimental data and KNO scaling relations.  The multiplicity of pions is required to be larger than one for $W<2$~GeV/$c^2$, because single pion production in this region is already taken into account (see Sec.~\ref{subsubsec:singlemeson}).

For coherent single $\pi$ production, we use the model developed by Rein and Sehgal~\cite{Rein:1983pf}. However, we only consider the neutral-current interaction, because the cross-section of charged-current coherent pion production was found to be very small in our neutrino energy region~\cite{Hasegawa:2005td}.

\subsubsection{Nuclear effects}
\label{subsubsec:nuclear}
For the interaction of neutrinos with nucleons inside the nucleus, the Fermi motion of nucleons and the Pauli exclusion principle are taken into account.  The momentum distribution of the target nucleon is assumed to be flat up to a fixed Fermi surface momentum of 225~MeV/$c$ for carbon.
%and oxygen and 250 MeV/$c$ for iron.  
The nuclear potential is set to 27~MeV.
%for carbon and oxygen and 32~MeV for iron.
The effect of Pauli blocking in the resonance process is implemented by suppressing resonance interactions in which the momentum of the outgoing nucleon is less than the Fermi surface momentum.

The interactions of $\pi$, $K$, $\eta$, and nucleons inside the target nucleus are simulated using a cascade model.  For pions, inelastic scattering, charge exchange, and absorption are considered.  For nucleons, elastic scattering and delta production are considered.
It is also possible for a delta resonance to be absorbed by the nucleus, meaning there is no pion in the final state.
% In our simulation, this occurs for 20\% of the deltas produced.

Nuclear interactions have a substantial impact on the final state particles.  Our simulation predicts that in 37\% of CC single charged pion production interactions in SciBar, the $\pi^{+}$ does not escape the nucleus: the delta resonance is absorbed in the nucleus 18\% percent of the time, the $\pi^{+}$ is absorbed 15\% of the time, and charge exchange occurs 4\% of the time.

Further details of the neutrino interaction simulation can be found in~\cite{Ahn:2006zz,Hayato:2002sd}.

\subsection{SciBar Detector Response}
\label{subsec:simulation_detector}
% Editors: Ana, Lisa

% Relevant features of GEANT detector simulation.

GEANT3~\cite{Brun:1987ma} is used to track 
the particles in the SciBar detector. The GCALOR program library~\cite{Zeitnitz:1994bs} is used to simulate the interactions of hadronic particles with the detector material.
The energy loss of a particle in each single strip is simulated by GEANT, and this value is adjusted according to the detector simulation.  The effect of scintillator quenching is simulated for protons, using the measured value for Birk's constant~\cite{birks} in SciBar, 0.0208 $\pm$ 0.0023 cm/MeV~\cite{Hasegawa-thesis}.  The attenuation of light in the WLS fiber is taken into account using the measured attenuation length for each channel, which is approximately 350 cm on average.  For each hit, crosstalk among nearby channels is simulated.  After these effects are simulated, the energy deposition in MeV is converted to number of photoelectrons (p.e.) using the light-yield calibration constant which is measured for each strip with cosmic muons.  The number of photoelectrons is then smeared by Poisson statistics.  Finally, the PMT single photoelectron resolution of 40\% is taken into account.  To simulate the digitization of the signal, the energy deposition in p.e. is converted to ADC counts.  Electronics noise and the response of the VA shaping are taken into account.

The simulated time of energy deposition is adjusted for the travel time of the light in the WLS fiber; the velocity of light in the fiber is approximately 16 cm/ns.  The simulated time is also smeared by the timing resolution.

\section{Analysis}
\label{sec:analysis1}
% Editors: Ana, Lisa

The goal of this analysis is to measure the cross section for CC single charged pion production in the resonance region, $\nu_{\mu} N \rightarrow \mu^{-} N \pi^{+}$ (CC1$\pi^{+}$), relative to the cross section of CCQE.  The CC1$\pi^+$ cross section is normalized to the CCQE cross section in this analysis to reduce the impact of neutrino flux uncertainties.  The dominant contribution to CC1$\pi^{+}$ is the $\nu_{\mu} p \rightarrow \mu^{-} p \pi^{+}$ (CCp$\pi^{+}$) mode, with a small contribution from the $\nu_{\mu} n \rightarrow \mu^{-} n \pi^{+}$ (CCn$\pi^{+}$) mode.

To make our measurement, we bin the data and perform a maximum likelihood fit to determine the cross sections of CC1$\pi^{+}$ and CCQE relative to the MC predictions.  From the MC prediction and the results of the fit, the observed CC1$\pi^{+}$ to CCQE cross section ratio can be extracted.  The data are divided into four samples that differ in their relative contributions from CCQE, CC1$\pi^{+}$ and other neutrino interactions.  The data in these four samples is binned using basic muon kinematic variables.

Note that we are concerned only with single pion production in the resonance region; in other words, our definition of CC1$\pi^{+}$ only includes neutrino interactions with $W<2$~GeV/$c^2$ that produce a single $\pi^{+}$.  We enforce this definition in the MC; the fit adjusts the rate of this interaction, among other things. The best fit predicts the CC1$\pi^{+}$ interaction rate in the data sample.  We do not attempt to identify CC1$\pi^{+}$ events in the data on an event-by-event basis.

We report the cross section for the interaction of the neutrino with the nucleon inside the nucleus, correcting for hadronic final state interactions.  For example, an event in which a $\pi^{+}$ produced via resonance is absorbed in the nucleus is included in our definition of CC1$\pi^{+}$.  We do not report the cross section after final state interactions because it is difficult to find pion tracks in the final state, making the experimental signature based on the full final state topology (rather than on muon kinematics, as discussed below) difficult to accomplish.  In addition, a measurement including final state interactions cannot easily be compared to measurements made with nuclear targets other than carbon.

We measure both the total cross section ratio and the energy-dependent cross section ratio.  The energy bins used in the energy-dependent measurement are 0-1.35, 1.35-1.72, 1.72-2.22, and $>$2.22 in units of GeV, denoted by indices $k$ = 0, 1, 2, and 3, respectively, in the text.  These bins are chosen so that the statistical uncertainty in the cross section ratio in each bin is similar. According to our simulation, the average true neutrino energy of all interactions that are either CC1$\pi^{+}$ or CCQE is 0.97, 1.52, 1.94 and 2.65 GeV for the four energy bins, respectively.

\subsection{Event Selection and Classification}
\label{subsec:analysis1_selection}
% Editors: Lisa

% MRD sample, classify events based on track multiplicity at
% vertex, $\Delta\theta_p$, reconstructed neutrino energy bins.
% Resulting number of events in data, and plots related to event 
% classification variables.

The fiducial volume of SciBar is defined to be 2.6 m in both the x and y directions and 1.35 m in the beam direction, for a total fiducial mass of 9.38 tons.  To select charged-current events, we identify the muon produced in the interaction.  We search for tracks starting in the fiducial volume of SciBar and in time with the beam that are matched with a track in the MRD or with hits in the first layer of the MRD.  For track matching, the MRD track is required to start in the first layer of the MRD and stop inside the MRD.  The distance between the extrapolation of the SciBar track and the actual MRD track starting point must be less than 20 cm, and the angular difference of the tracks must be less than 0.5 radians in both projections.  For matching a SciBar track to first layer MRD hits, the distance between the extrapolation of the SciBar track and the hits in the MRD must be less than 20 cm.  This SciBar-MRD matched track is identified as the muon for the event.  If there is more than one SciBar-MRD matched track, the most energetic one is defined as the muon track.  The MRD matching requirement imposes a muon momentum ($p_{\mu}$) threshold of 450 MeV/$c$.
The sample of events in which a SciBar-MRD track is found is our CC sample.  We veto events with hits in the first layer of SciBar to eliminate events due to incoming particles produced by neutrino interactions in the other near detectors.  According to MC simulation, 96\% of the events in the CC sample are true CC interactions.  In addition, approximately 70\% of true CC events that produce a muon with momentum of at least 450 MeV/$c$ are successfully reconstructed in the CC sample, the main inefficiency being due to unsuccessful muon track reconstruction in the SciBar detector.

The 3D angle of the muon with respect to the $z$-axis ($\theta_{\mu}$) can be calculated using the slopes of the track in each 2D projection.
The energy of the muon is reconstructed by the range and expected energy deposit per unit length in SciBar, the EC, and the MRD.  In Equation \ref{emu}, 
\begin{align}
\label{emu}
 E_{\mu} &= E_{\mu}^{SciBar} + E_{\mu}^{EC} + E_{\mu}^{MRD}\notag\\
 &= L_{SciBar}(\frac{dE}{dx})_{SciBar} + \frac{dE_{0,EC}}{\cos\theta_\mu} + E_{\mu}^{MRD},
\end{align}
$E_{\mu}^{SciBar}$, $E_{\mu}^{EC}$, and $E_{\mu}^{MRD}$ are the energy deposited in each detector. $L_{SciBar}$ is the muon's range through SciBar; $\displaystyle(\frac{dE}{dx})_{SciBar}$ is set to 2.10 MeV/cm. $dE_{0,EC}$, set to 90 MeV, is the most probable value for the energy deposited in the EC for a muon crossing the EC planes perpendicularly, as estimated from the MC simulation~\cite{Brun:1987ma}.
$E_{\mu}^{MRD}$ is calculated from a range to energy lookup table based on GEANT3~\cite{Brun:1987ma} MC and includes the muon's mass.  The average muon momentum and muon angle resolutions are 90 MeV/$c$ and 1.4 degrees, respectively.

\begin{figure}[htp]
  \begin{center}
    \includegraphics[width=0.45\textwidth]{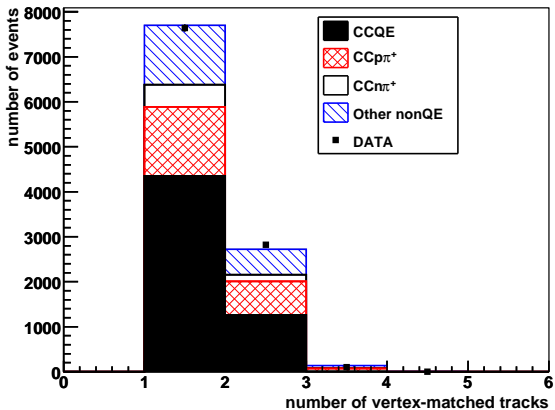}
  \end{center}
  \caption{\label{fig:NtrackAtVtx} (Color online) Number of vertex-matched tracks.  Most events are 1-track, i.e. only the SciBar-MRD track.  $\chi^2$/d.o.f = 3.46/4 considering the systematic errors discussed in Section \ref{sec:systematics}.}
\end{figure}

The upstream endpoint of the SciBar-MRD track is the reconstructed event vertex.  We apply a vertex matching cut to other tracks in each event.  We search for tracks starting in the fiducial volume that have an edge that is no more than 4.8 cm from the reconstructed event vertex and are in coincidence with the SciBar-MRD track within 100 ns.  The 4.8 cm cut corresponds to 3$\sigma$ in vertex resolution in the $z$ dimension and more than 5$\sigma$ in the $x$ and $y$ dimensions.  Figure \ref{fig:NtrackAtVtx} shows the distribution of vertex-matched tracks for data and MC.  For the MC, the contributions from CCQE, CCp$\pi^{+}$, CCn$\pi^{+}$, and other nonQE interactions are shown separately.

\begin{figure}[htp]
  \begin{center}
    \includegraphics[width=0.45\textwidth]{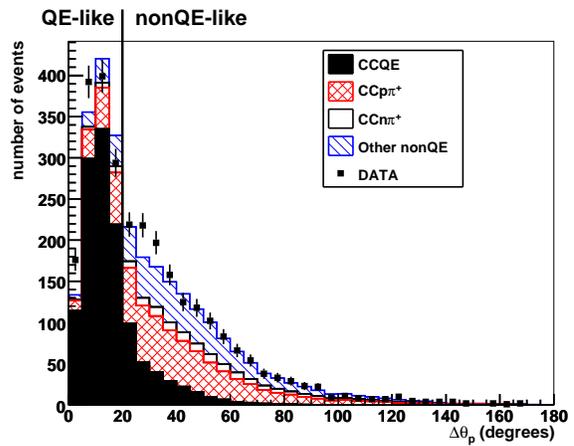}
  \end{center}
  \caption{\label{fig:Dtheta} (Color online) $\Delta \theta_{p}$, angle between expected proton track assuming CCQE and the observed second track for the 2-track sample.  $\chi^2$/d.o.f = 32.58/30 considering the systematic errors discussed in Section \ref{sec:systematics}.}
\end{figure}

Since the CCQE interaction is a two-body interaction, the direction of the proton can be calculated given the momentum and direction of the muon.  For 2-track events, we define an angle called $\Delta \theta_{p}$ which is the angle between the expected proton track (calculated assuming the event was CCQE) and the observed second track.  Figure \ref{fig:Dtheta} shows the distribution of $\Delta\theta_{p}$ for data and MC.  For the MC, the contributions from CCQE, CCp$\pi^{+}$, CCn$\pi^{+}$, and other nonQE interactions are shown separately.  There is some apparent discrepancy between data and MC in Figure \ref{fig:Dtheta}; however, when systematic uncertainties are taken into account (see Section \ref{sec:systematics}), the data and MC are consistent.  2-track events with $\Delta \theta_{p}$ less than 20 degrees are considered QE-like, and all other 2-track events are considered nonQE-like. The cut value of 20 degrees is chosen to maximize the purity squared times efficiency of selecting CC1$\pi^{+}$ events in the 2-track nonQE-like sample.

\begin{figure}[htp]
  \begin{center}
    \includegraphics[width=0.45\textwidth]{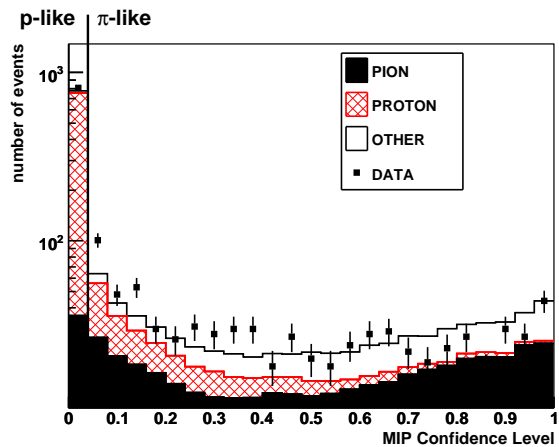}
  \end{center}
  \caption{\label{fig:MuCL} (Color online) MIP Confidence Level for the second track in the 2-track nonQE sample in log scale.  $\chi^2$/d.o.f = 6.36/20 considering the systematic errors discussed in Section \ref{sec:systematics}.}
\end{figure}

SciBar has the capability to distinguish protons from muons and pions using $dE/dx$.  The MIP confidence level (MIPCL) is related to the probability that a particle is a minimum ionizing particle based on the energy deposition.  The confidence level per layer is the fraction of events in the muon $dE/dx$ distribution (obtained from cosmic muons) with larger energy deposition than what is observed in that layer.  The total confidence level, MIPCL, is obtained by assuming the confidence level at each layer is independent and calculating the combined probability.  This variable is considered for vertex-matched tracks other than the SciBar-MRD track.  Tracks with MIPCL less than 0.04 are considered proton-like and all other tracks are considered pion-like.  A MIPCL cut at 0.04 maximizes the purity squared times efficiency of selecting CC1$\pi^{+}$ events in the 2-track nonQE-like pion-like sample.
Figure \ref{fig:MuCL} shows the distribution of MIPCL for the second track in the 2-track nonQE-like sample for data and MC.  For the MC, the distributions for protons, pions, and other particles are shown separately.

For this analysis, we consider only 1- and 2-track events.  2-track events are classified as QE- or nonQE-like based on $\Delta\theta_{p}$.  2-track nonQE-like events are classified as pion- or proton-like based on the MIPCL of the second track.  Thus there are four samples of events: 1-track, 2-track QE, 2-track nonQE pion, and 2-track nonQE proton.  Table \ref{tab:EventSamples} shows the number of data events in each sample and the composition as predicted by our nominal MC simulation.  Contamination from backgrounds that are not neutrino-induced is negligible.  The combined efficiency for selecting CC1$\pi^{+}$ events in these samples is 44\%.

\begin{table}[htp]
\caption{Event Samples}
\label{tab:EventSamples}
\centering
\begin{tabular}{lccccc}
\hline
\hline
Sample&Data&CCQE&CCp$\pi^{+}$&CCn$\pi^{+}$&Other\\
&(\# events)&&&&nonQE\\
\hline
\hline
\textbf{1-trk}&7638&57\%&20\%&6\%&17\%\\
\textbf{2-trk QE}&1261&78\%&13\%&1\%&8\%\\
\textbf{2-trk nonQE $\pi$}&750&6\%&41\%&15\%&38\%\\
\textbf{2-trk nonQE p}&811&32\%&38\%&3\%&27\%\\
\textbf{total selected}&10400&54\%&22\%&6\%&18\%\\
\hline
\hline
\end{tabular}
\end{table}

\subsection{Cross Section Ratio Extraction}
\label{subsec:analysis1_corrections}

\begin{figure}[htp]
  \begin{center}
    \includegraphics[width=0.4\textwidth]{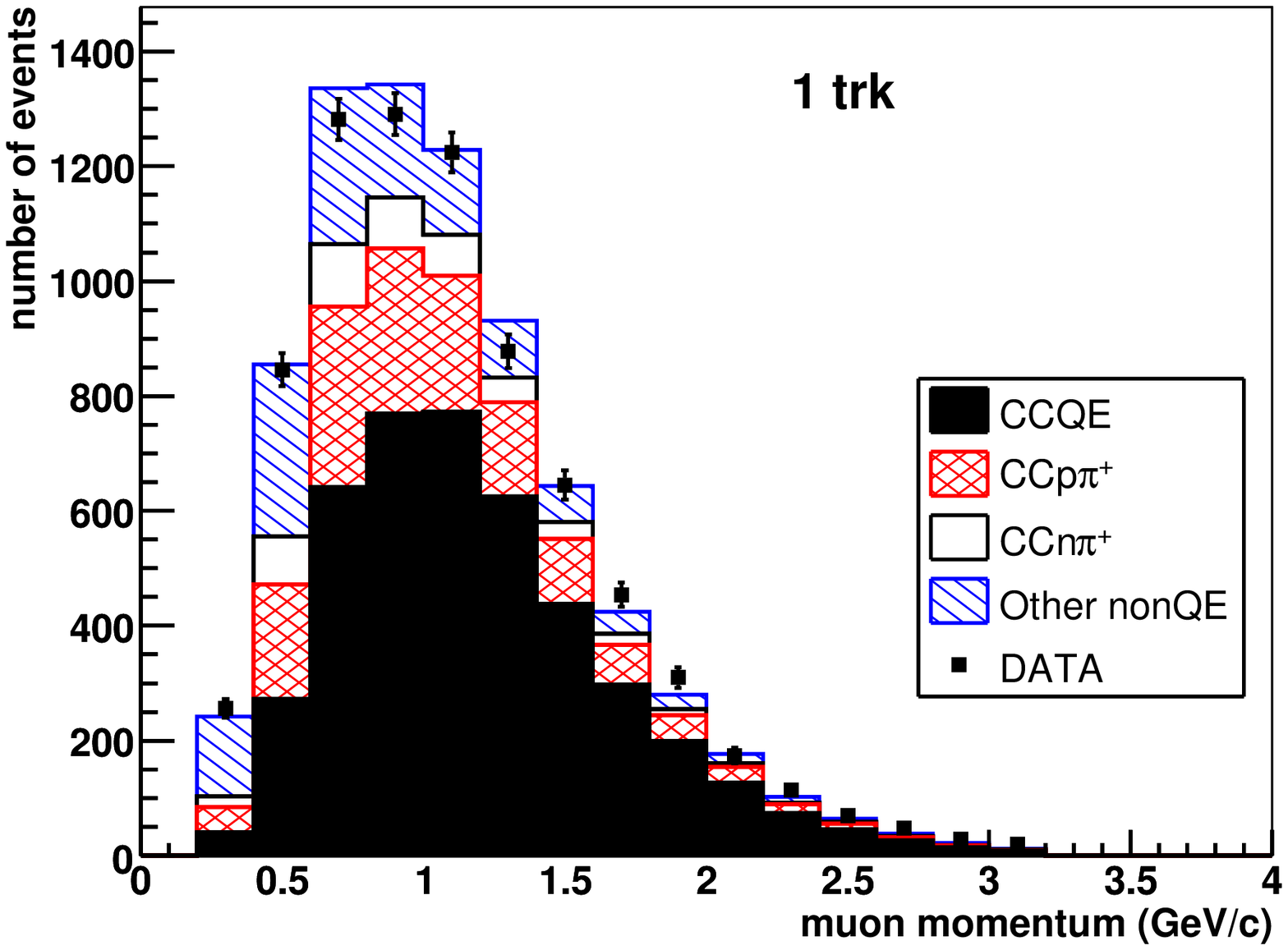}
    \includegraphics[width=0.4\textwidth]{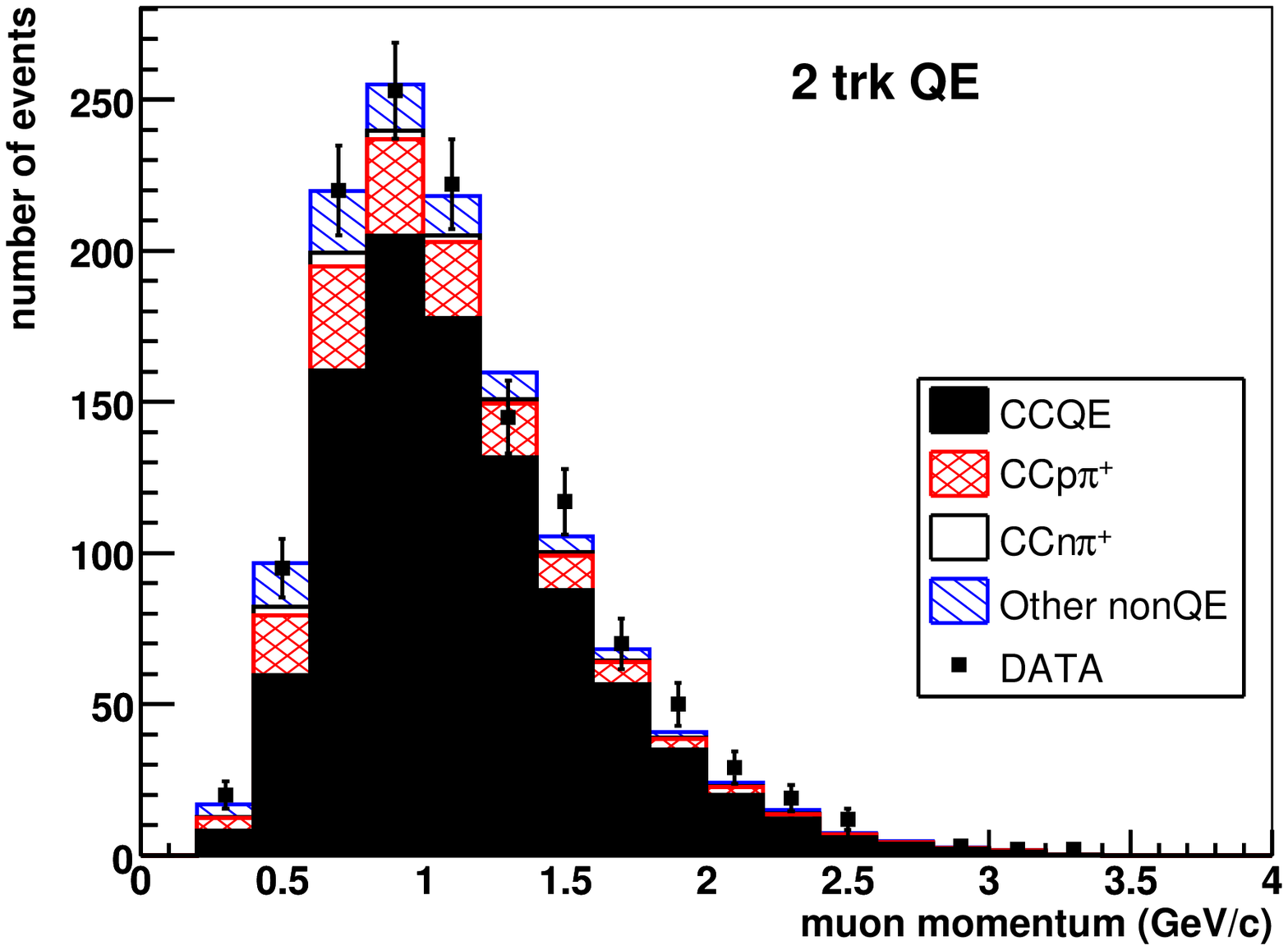}
    \includegraphics[width=0.4\textwidth]{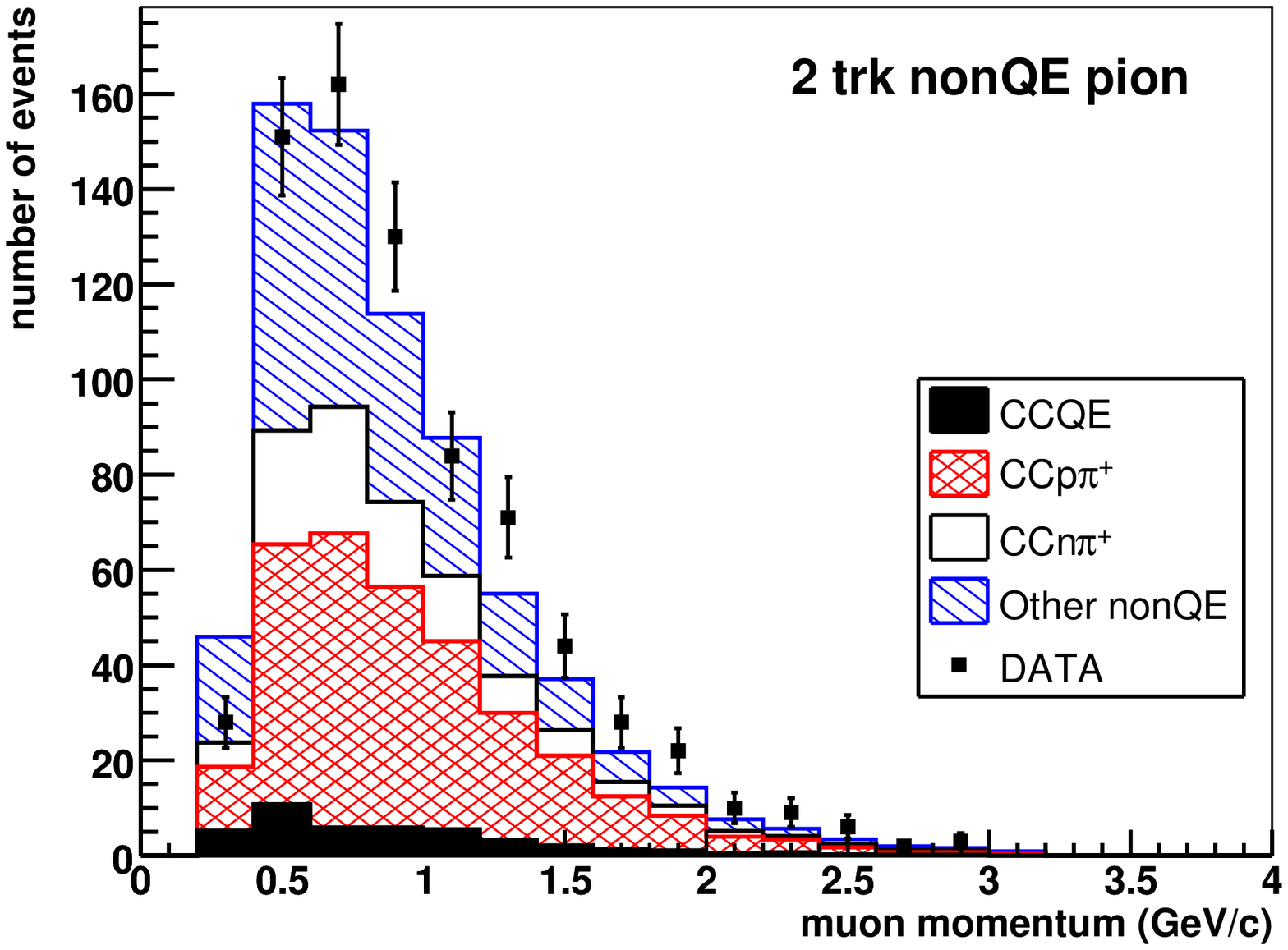}
    \includegraphics[width=0.4\textwidth]{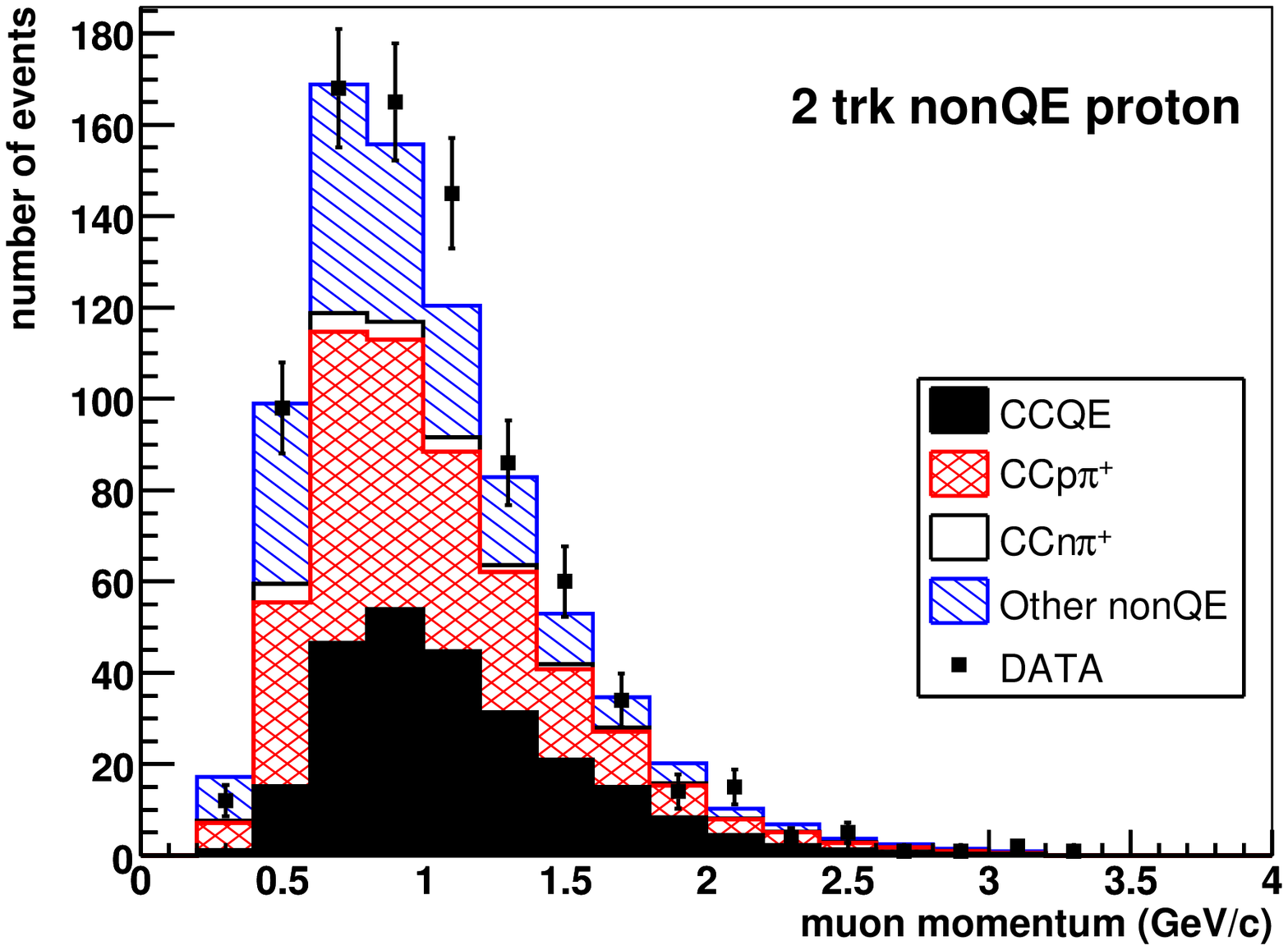}
  \end{center}
  \caption{\label{fig:pmu} (Color online) Muon momentum distributions for each of the four data samples used in this analysis.}
\end{figure}

\begin{figure}[htp]
  \begin{center}
    \includegraphics[width=0.4\textwidth]{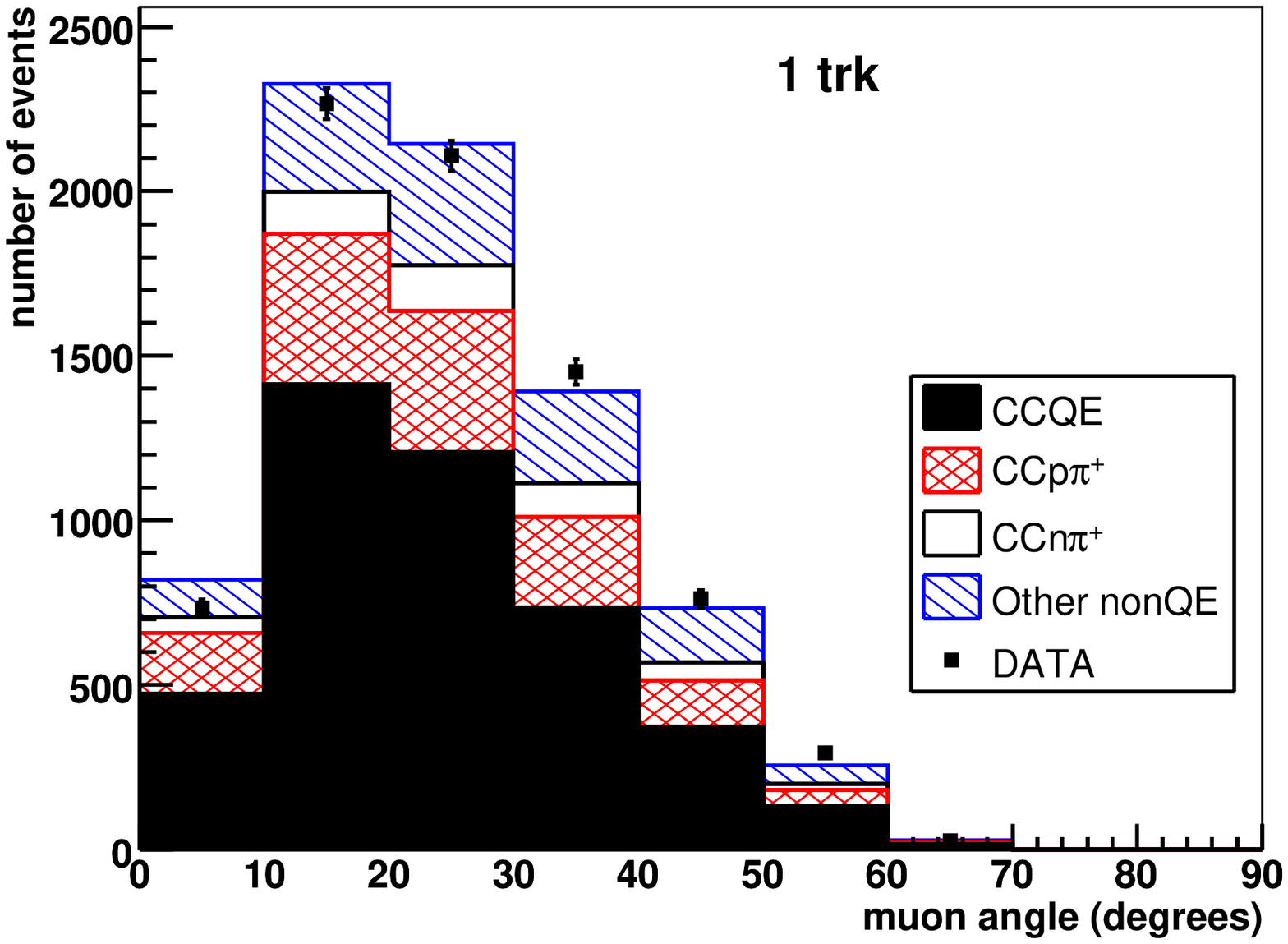}
    \includegraphics[width=0.4\textwidth]{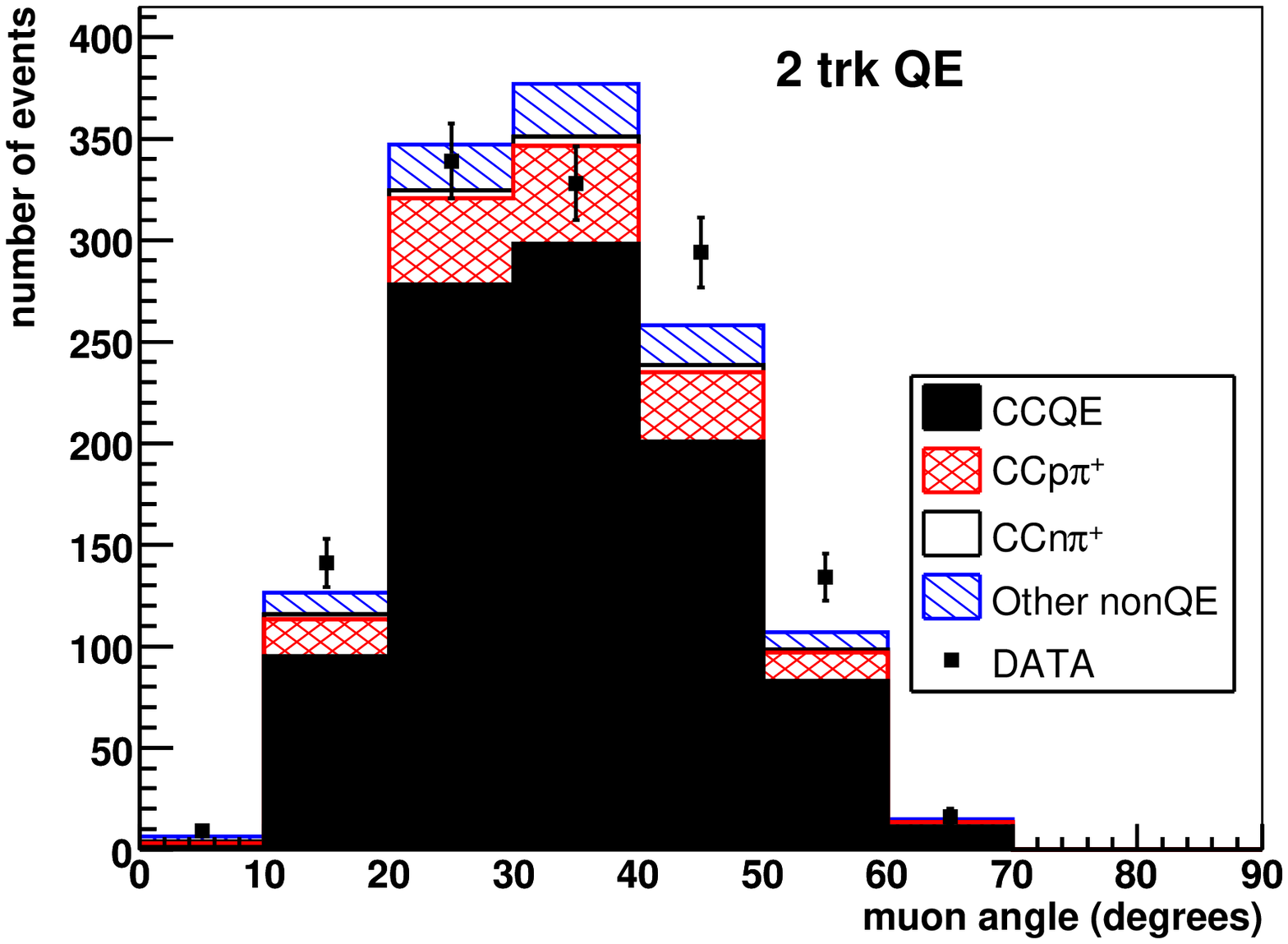}
    \includegraphics[width=0.4\textwidth]{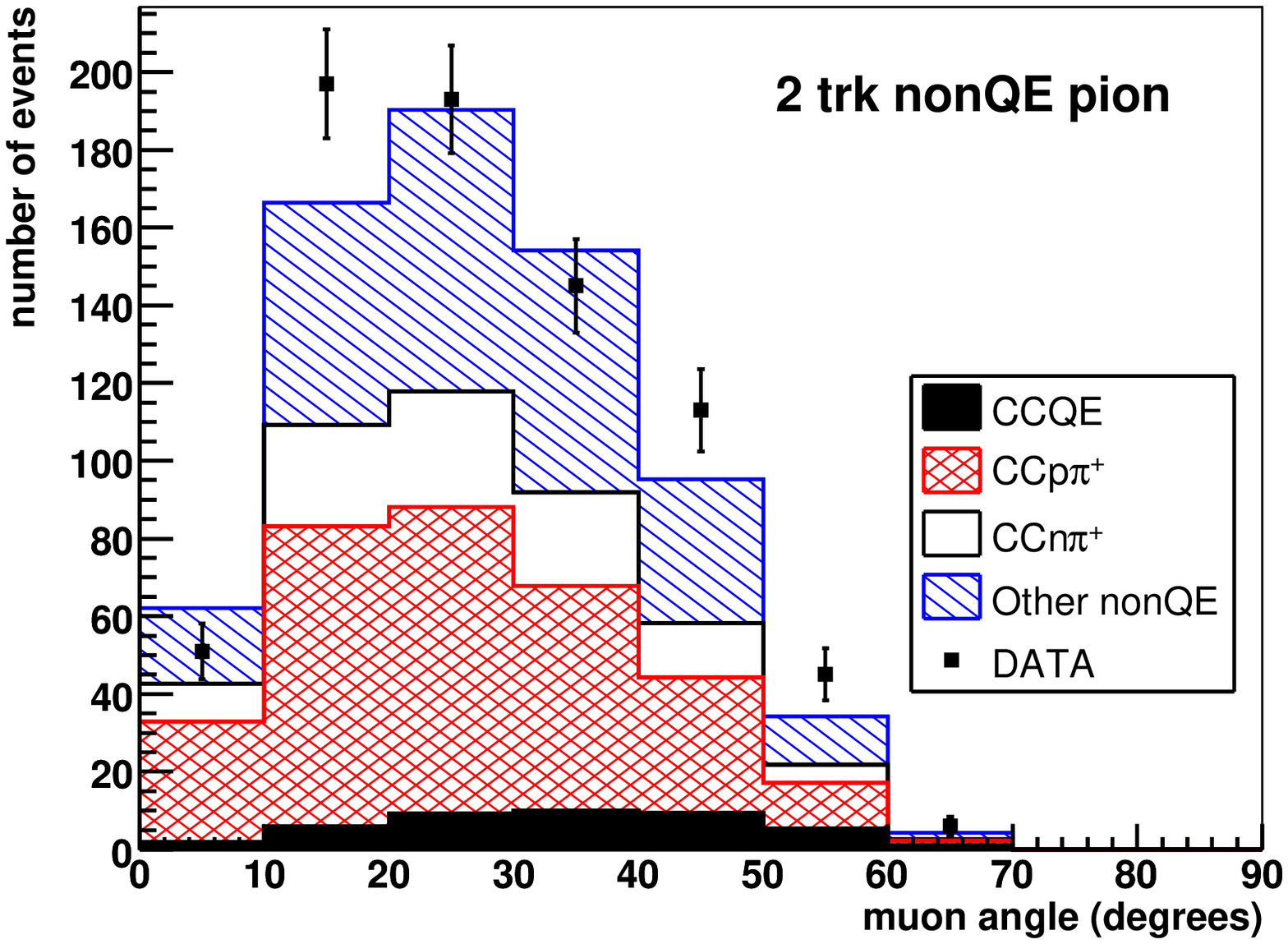}
    \includegraphics[width=0.4\textwidth]{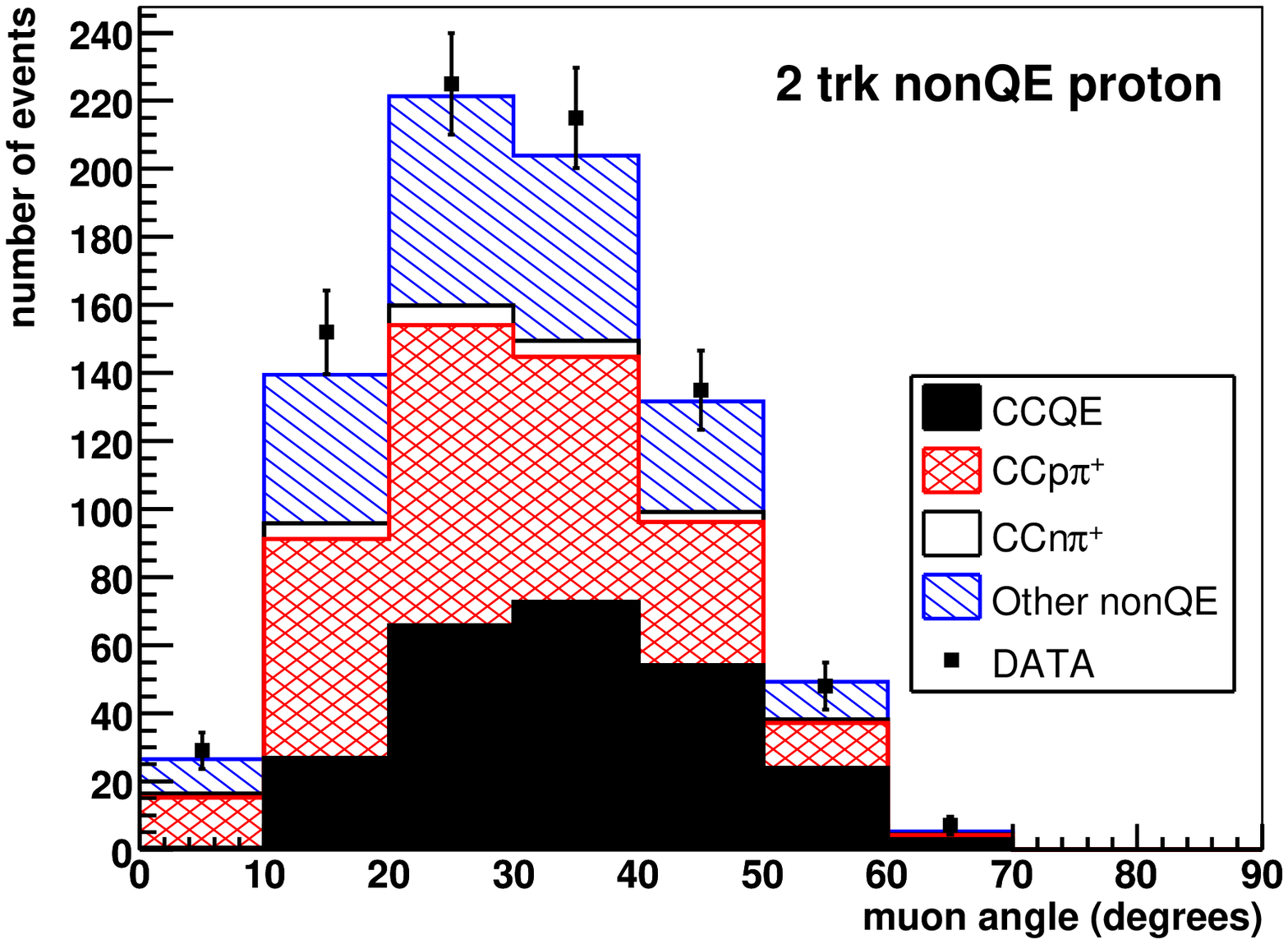}
  \end{center}
  \caption{\label{fig:qmu} (Color online) Muon angle distributions for each of the four data samples used in this analysis.}
\end{figure}

The MC events in the four samples are divided based on the interaction type. The interaction types considered are CCQE, CC1$\pi^{+}$, and other nonQE, where `other' refers to nonQE interactions other than CC1$\pi^{+}$. In addition, MC events are divided based on true neutrino energy so that an energy-dependent measurement of CC1$\pi^{+}$ interactions can be performed. Data and MC are then binned in $p_{\mu}$ bins of size 0.2 GeV/$c$ and $\theta_{\mu}$ bins of size 10 degrees.  The momentum $p_{\mu}$ ranges from 0 to 4 GeV/$c$, and the angle $\theta_{\mu}$ ranges from 0 to 90 degrees.  Thus, there are a total of 180 bins in the $p_{\mu}$ vs. $\theta_{\mu}$ distributions.  Figure \ref{fig:pmu} shows the muon momentum ($p_{\mu}$) distribution for each sample for data and MC, and Figure \ref{fig:qmu} shows the muon angle ($\theta_{\mu}$) distribution for each sample for data and MC.  For the MC, the contributions from CCQE, CCp$\pi^{+}$, CCn$\pi^{+}$, and other nonQE interactions are shown separately.

Only bins that are expected to have enough statistics are used in the fit.  The requirement for a bin to be used in the fit is that the nominal MC (normalized to data) predicts at least three events in that bin.  Of the 180 bins in each $p_{\mu}$ vs. $\theta_{\mu}$ distribution, 66, 42, 43, and 37 bins satisfy this requirement in the 1-track, 2-track QE, 2-track nonQE pion, and 2-track nonQE proton samples, respectively.

We use the method of maximum likelihood for the fit.  For Poisson statistics, maximizing the likelihood ratio is equivalent to minimizing the quantity~\cite{Yao:2006px}
\begin{equation}
\label{eqn:F}
F = 2\sum_{i,j} \left[ N^{exp}_{i,j} - N^{obs}_{i,j} + N^{obs}_{i,j}\ln\frac{N^{obs}_{i,j}}{N^{exp}_{i,j}} \right],
\end{equation}
where $N^{exp}_{i,j}$ and $N^{obs}_{i,j}$ are the number of expected events and number of observed events in $p_{\mu}$ vs. $\theta_{\mu}$ bin $i$ ($i$=1,...,180) for sample $j$ ($j$=1,...,4), respectively.

The number of expected events in a given bin is a function of the nominal MC and the fitting parameters, as shown in Equation \ref{eqn:nexp},
\begin{align}
\label{eqn:nexp}
N^{exp}_{i,j} = ~ &\alpha [N^{CCQE}_{i,j} + \sum_{k} (R^{CC1\pi^{+}}_{k} N^{CC1\pi^{+}}_{i,j,k}) ~~ + \notag\\ 
&R^{OtherNonQE} N^{OtherNonQE}_{i,j}],
\end{align}
where $k$ represents a bin of true neutrino energy, $N^{CCQE}_{i,j}$ ($N^{OtherNonQE}_{i,j}$) is the number of CCQE (other nonQE) events in bin $i$ in sample $j$ for the nominal MC, and $N^{CC1\pi^{+}}_{i,j,k}$ is the number of CC1$\pi^{+}$ events in bin $i$ in sample $j$ for true neutrino energy bin $k$ in the nominal MC.

The goal is to simultaneously fit for the contributions of CCQE, CC1$\pi^{+}$, and other nonQE to the data distributions in each sample. In Equation \ref{eqn:nexp}, the fitting parameters are $P_{sc}$, $R^{CC1\pi^{+}}_{k}$, and $R^{OtherNonQE}$.  $R^{CC1\pi^{+}}_{k}$ scales the fraction of CC1$\pi^{+}$ events depending on the true energy.  (For the energy-independent fit, all CC1$\pi^{+}$ events are scaled the same regardless of the true energy, and the subscript $k$ is not used.) $R^{OtherNonQE}$ scales the overall fraction of other nonQE events in the sample.  The nominal values of $R^{OtherNonQE}$ and  $R^{CC1\pi^{+}}_{k}$ are 1.  All three interaction types are scaled by the data to MC normalization, $\alpha$.  $\alpha$ is not constant; it is adjusted at each iteration of the fit so that $\displaystyle \sum_{i,j} N^{exp}_{i,j} \equiv \sum_{i,j} N^{obs}_{i,j}$, i.e. the total number of expected events in the four samples is fixed to the observed number.

For the energy-dependent fit, the scaling for the CC1$\pi^{+}$ fraction is energy-dependent. However, this is not true for CCQE.  We estimate the overall fraction of CCQE events, and we fix the energy dependence of the CCQE cross section to the MC prediction because it has been accurately measured by previous experiments (\cite{Barish:1977qk,Bonetti:1977cs,ggm,Belikov:1983kg}).  The uncertainty in the $Q^2$ dependence of the CCQE cross section is considered as a systematic error.  Similarly, we estimate the overall fraction of other nonQE events and assume that the energy dependence of the other nonQE cross section is correctly described by our MC simulation.

The muon momentum scale, $P_{sc}$, does not appear explicitly in Equation \ref{eqn:nexp}, but it is a free parameter in the fit.  The purpose of $P_{sc}$ is to allow shrinking or stretching of the distributions along the $p_{\mu}$ axis; in this way we account for a systematic difference in the energy scale between data and MC.  The first step in each iteration of the minimization is rescaling the $p_{\mu}$ vs. $\theta_{\mu}$ distributions for the nominal MC by $P_{sc}$.  This means that $N^{CCQE}_{i,j}$, $N^{OtherNonQE}_{i,j}$, and $N^{CC1\pi^{+}}_{i,j,k}$ change slightly from their nominal MC predictions due to shifting of events among bins.  The nominal value of $P_{sc}$ is 1.

There is an additional term added to the minimization function shown in Equation \ref{eqn:F}.  The systematic error in the muon momentum scale is estimated to be 2.7\%, dominated by uncertainties of muon energy reconstruction in the MRD (see Section \ref{subsubsec:MRD}).  $P_{sc}$ is a free parameter, but its fit value is constrained by the estimated systematic error.  The following term is added to the minimization function to accomplish this:
\begin{equation}
F_{Psc} = \frac{(P_{sc}-1)^{2}}{(0.027)^{2}}.
\end{equation}

The CC1$\pi^{+}$ contribution to the selected sample for true neutrino energy in bin $k$ is changed from the nominal MC prediction by a factor of $R^{CC1\pi^{+}}_{k} \times (\alpha/\alpha_{nominal})$, where $\alpha$ is the normalization at best fit and $\alpha_{nominal}$ is the normalization for the nominal MC.  The other nonQE contribution to the selected sample is changed by a factor of $R^{OtherNonQE} \times (\alpha/\alpha_{nominal})$, and the CCQE contribution to the selected sample is changed by a factor of $\alpha/\alpha_{nominal}$.

Tables \ref{tab:fit_1enubin} and \ref{tab:fit_4enubins} show the best fit values of $P_{sc}$, $R^{CC1\pi^{+}}_{k}$, $R^{OtherNonQE}$, and the normalization $\alpha$ relative to the nominal normalization for the energy-independent and energy-dependent fits, respectively.  The best fit value of $P_{sc}$, though significantly different from 1, is consistent with previous K2K measurements.

\begin{table}[htp]
\caption{Best Fit Parameter Values (energy-independent CC1$\pi^{+}$)}
\label{tab:fit_1enubin}
\centering
\begin{tabular}{c|c}
\hline
\hline
\textbf{Parameter}&\textbf{Best Fit Value}\\
\hline
$P_{sc}$&0.974$\pm$0.004\\
$R^{CC1\pi^{+}}$&0.992$\pm$0.116\\
$R^{OtherNonQE}$&1.309$\pm$0.119\\
$\alpha/\alpha_{nominal}$&0.951$\pm$0.043\\
\hline
\hline
\end{tabular}
\end{table}

\begin{table}[htp]
\caption{Best Fit Parameter Values (energy-dependent CC1$\pi^{+}$)}
\label{tab:fit_4enubins}
\centering
\begin{tabular}{c|c}
\hline
\hline
\textbf{Parameter}&\textbf{Best Fit Value}\\
\hline
$P_{sc}$&0.977$\pm$0.005\\
$R^{CC1\pi^{+}}_{0}$&0.750$\pm$0.208\\
$R^{CC1\pi^{+}}_{1}$&1.106$\pm$0.180\\
$R^{CC1\pi^{+}}_{2}$&0.900$\pm$0.191\\
$R^{CC1\pi^{+}}_{3}$&1.105$\pm$0.246\\
$R^{OtherNonQE}$&1.379$\pm$0.136\\
$\alpha/\alpha_{nominal}$&0.949$\pm$0.047\\
\hline
\hline
\end{tabular}
\end{table}

The minimum of the fitting function (Equation \ref{eqn:F}) follows a $\chi^2$ distribution and can be used to estimate the goodness of the fit ~\cite{Yao:2006px}.  The $\chi^2$/d.o.f before the fit is 283/187 = 1.51.  The $\chi^2$/d.o.f. at best fit is 229/185 = 1.24 and 228/182 = 1.25 for the energy-independent and energy-dependent fits, respectively.  (The $\chi^2$ including all systematic uncertainties is given in Section \ref{subsec:discusssyst}).  For the energy-independent fit, the predictions for the number of CCQE and CC1$\pi^{+}$ events in the sample are consistent with the nominal MC within fitting errors.  For the energy-dependent fit, the prediction for the number of CCQE events is consistent with the nominal MC within fitting errors, and the predictions for the number of CC1$\pi^{+}$ events in each neutrino energy bin are consistent with the nominal MC within fitting errors except in the first neutrino energy bin.
In addition, we find that the ratio of other nonQE interactions to CCQE interactions in our data is larger than this ratio in the nominal MC by 31\% $\pm$ 12\% (38\% $\pm$ 14\%) for the energy-independent (energy-dependent) fit, where the uncertainty is due to fitting only and does not include systematic uncertainties.  
As a cross-check, we further imposed the condition $R_{k}^{CC1\pi^{+}} = R^{OthernonQE} \equiv R^{nQE}, (k=0,1,2,3)$ in the fit function given in Equation \ref{eqn:nexp} to extract the relative weighting of all CC nonQE events to CCQE events, as done in~\cite{Ahn:2006zz}. With this assumption, we obtain $\chi^2$/d.o.f = 231/185 = 1.25 and $R^{nQE} = 
1.151 \pm 0.057$, to be compared with the 1.194 $\pm$ 0.092 value quoted in~\cite{Ahn:2006zz}.

\begin{figure}[htp]
  \begin{center}
    \includegraphics[width=0.45\textwidth]{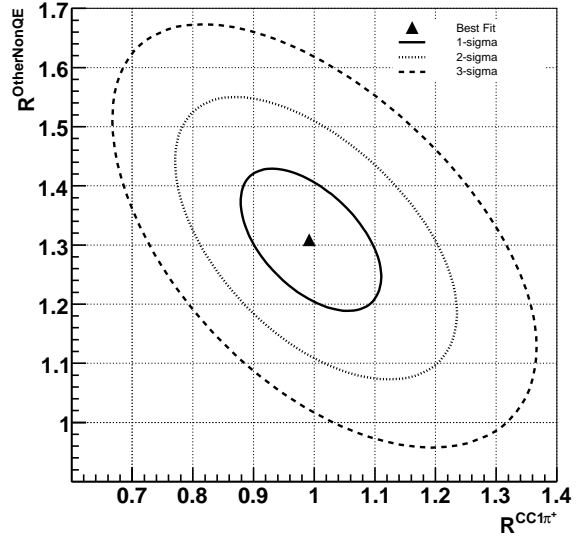}
  \end{center}
  \caption{\label{fig:Contour} 1-, 2-, and 3-sigma contours for parameters $R^{CC1\pi^{+}}$ and $R^{OtherNonQE}$ for the energy-independent fit.}
\end{figure}

In this analysis, we fully take into account correlations among fit parameters.  Figure \ref{fig:Contour} shows the best fit point and the 1-, 2-, and 3-sigma contour lines for parameters $R^{CC1\pi^{+}}$ and $R^{OtherNonQE}$ for the energy-independent fit.  The correlation between $R^{CC1\pi^{+}}$ and $R^{OtherNonQE}$ is -0.539; for the energy-dependent fit, the correlation of $R^{CC1\pi^{+}}_{0}$, $R^{CC1\pi^{+}}_{1}$, $R^{CC1\pi^{+}}_{2}$, and $R^{CC1\pi^{+}}_{3}$ with $R^{OtherNonQE}$ is -0.644, -0.158, -0.312, and -0.216, respectively.

The CC1$\pi^{+}$ to CCQE cross section ratio for neutrino energy bin $k$, $R_{k}$, can be calculated under the assumption that the efficiencies for detecting CCQE and CC1$\pi^{+}$ interactions in the selected sample are the same in data and MC.
\begin{align}
R_{k} &= \frac{\sigma^{CC1\pi^{+}}_{k}}{\sigma^{CCQE}_{k}}\notag\\
&= \frac{N^{CC1\pi^{+}}_{k}(data)}{N^{CC1\pi^{+}}_{k}(MC)} \times \frac{N^{CCQE}_{k}(MC)}{N^{CCQE}_{k}(data)} \times R_{MC,k},
\end{align}
where $N^{CC1\pi^{+}}_{k}$ and $N^{CCQE}_{k}$ are the numbers of CC1$\pi^{+}$ and CCQE events, respectively, in neutrino energy bin $k$ in the selected sample and $R_{MC,k}$ is the MC prediction for the cross section ratio in neutrino bin $k$.  The data values in the above equation are obtained from the best fit MC.

\section{Systematic Effects}
\label{sec:systematics}
% Editors: Ana, Lisa

For each systematic source,  a new set of $p_{\mu}$ vs. $\theta_{\mu}$ distributions are made with the altered MC.  The fitting is performed again with these new distributions, and a different result for the cross section ratio is obtained, $R_{k}'$.  The systematic error is defined as the difference of the new result from the nominal, $\Delta R_{k}(syst) = R_{k}' - R_{k}$.  The effects contributing to our systematic uncertainty are described below.

\subsection{Interaction Model and Neutrino Flux}
\label{subsec:systematics_nuint}
% Editors: Ana
The axial-vector mass for the CCQE interaction is set in the neutrino interaction model to 1.1 GeV/$c^2$. The error in this value is estimated to be about $\pm$0.1 GeV/$c^2$ in an analysis of data from the SciFi detector~\cite{Gran:2006jn}. We quote a systematic error on $M_A^{CCQE}$ varying it by $\pm$ 0.1 GeV/$c^2$. In this analysis, we use the data to infer the overall fraction of CCQE interactions in our CC inclusive sample. The energy dependence of the CCQE fraction is assumed to be consistent with the MC prediction.  When varying $M_A^{CCQE}$, the total cross section is re-weighted back to the nominal value, so that the overall fraction of CCQE does not change.  Thus we can consider the systematic error due to the cross section shape only.

The uncertainty in the Bodek and Yang correction to DIS events is considered by changing the correction parameter by $\pm$ 30\% ~\cite{Bodek:2002vp}.  This translates into an uncertainty in the $p_{\mu}$ vs. $\theta_{\mu}$ shape prediction for other nonQE events and therefore is a source of uncertainty in the other nonQE fraction determined by our fit.

The uncertainty in the measurement of the neutrino energy spectrum used to tune the MC simulation~\cite{Ahn:2006zz} is considered by changing the tuning parameters within their errors.  Correlations among the tuning parameters are taken into account appropriately.

\subsection{Nuclear Effects}
\label{subsec:systematics_nucleareffects}
% Editors: Ana

The effect of changing the cross sections of pion absorption, pion scattering, and proton rescattering in the nucleus is considered as a systematic error. In the momentum range of pions from $\Delta^{++}$ decay, the cross section measurement uncertainty for both pion absorption and pion scattering is approximately 30$\%$~\cite{Ingram:1982bn}; therefore the cross section for pion absorption and pion scattering are each changed by $\pm$ 30$\%$ to evaluate the systematic error. The uncertainty in the cross section of proton rescattering inside the nucleus is about 10$\%$~\cite{Jeon-thesis}, and so the cross section of proton rescattering is varied by $\pm$ 10$\%$ to evaluate the systematic error.

In NEUT, the maximum Fermi momentum of nucleons is set to 225 MeV/$c$ for carbon~\cite{Ahn:2006zz}.
The value should be approximately 221 $\pm$ 5 MeV/$c$ according to~\cite{Moniz:1971mt}. We calculate the systematic error due to this effect by changing the maximum Fermi momentum by $\pm$ 5 MeV/$c$.

\subsection{Detector Response and Track Reconstruction}
\label{subsec:systematics_detector}
% Editors: Lisa

The model used to simulate crosstalk has one parameter, the amount of crosstalk in neighboring channels.  For the nominal MC, this parameter is set to 3.25\%. This value is chosen by tuning hit distributions. To evaluate the systematic effect of the crosstalk model, the crosstalk parameter is changed by its systematic error of 0.25\%, i.e. to 3.0\% and 3.5\%.  The model is adjusted simultaneously for both crosstalk simulation in the MC and the crosstalk correction in data and MC.

The single photoelectron resolution is nominally set at 40\%.  This value is chosen by tuning the simulated $dE/dx$ per plane for muons to match the observed values. The systematic error in this value was estimated to be 10\% in the tuning process.  Thus the resolution is changed to 30\% and 50\% to evaluate the uncertainty.

The model for scintillator quenching relies on Birk's constant, which is measured in SciBar to be 0.0208 $\pm$ 0.0023~\cite{Hasegawa-thesis}.  The constant is changed within its error to evaluate the systematic uncertainty.

A (software) hit threshold is nominally set at 2.0 photoelectrons (p.e.) in both data and MC to eliminate hits from noise in data that are not fully simulated. The estimated uncertainty in the photoelectron yield for a single hit is 15\%.  Thus, to evaluate the systematic error due to the threshold, we increase the threshold in MC by 15\% and assume that the change due to decreasing the threshold is the same magnitude.  We avoid decreasing the threshold, as a lower threshold would be in the region of data-MC discrepancy.

The difference in angular resolution between data and MC is considered as a systematic error.  The track fitting algorithm supplies the slopes of a track in the $x$-$z$ and $y$-$z$ planes, denoted $t_{x}$ and $t_{y}$, respectively.  To determine the angular resolution, we select good muon tracks.  Each track is divided into two halves, and then each half is fitted to get the $t_{x}$ value.  The difference between these two $t_{x}$ values is called $\sigma_{tx}$.  (The procedure is the same for $t_{y}$).  Both the data and MC distributions of $\sigma_{t}$ are fitted with Gaussians to get the resolution in data and MC, which are 30.91 mrad and 29.65 mrad, respectively.  We evaluate the systematic error due to difference in angular resolution by smearing $t_{x}$ and $t_{y}$ in the MC event-by-event by the difference of the resolutions, $\sqrt{30.91^2 - 29.65^2} = 8.74$ mrad.

\subsection{Discussion of Systematic Errors}
\label{subsec:discusssyst}
Tables \ref{tab:systerr_eind} and \ref{tab:systerr_edep} summarize the uncertainties for the energy-independent and energy-dependent ratio measurements, respectively.  In some cases, the uncertainty goes in the same direction for positive and negative variations of a certain effect.  In these cases, only the larger uncertainty is included in the total.  The systematic uncertainties that have the largest effect on the measurements presented in this paper are uncertainties in modeling final state interactions of pions and protons in the nucleus and uncertainty in modeling momentum transfer in CCQE interactions.  For the cross section ratio measurements as a function of neutrino energy, uncertainties in the neutrino spectrum measured at the near detector are also significant.

For the event selection variables, number of tracks (Figure \ref{fig:NtrackAtVtx}), $\Delta\theta_{p}$ (Figure \ref{fig:Dtheta}), and MIPCL (Figure \ref{fig:MuCL}), data and MC are consistent within the systematic uncertainties due to the sources already discussed.  Therefore we do not include additional systematic uncertainty due to the event selection variables, since this uncertainty is covered by what is given in Tables \ref{tab:systerr_eind} and \ref{tab:systerr_edep}.

To evaluate the goodness of fit considering systematics, we re-weight all the MC using the best fit parameters from the cross section analysis (Tables \ref{tab:fit_1enubin} and \ref{tab:fit_4enubins}) and add pull terms for each systematic effect to the function in Equation \ref{eqn:F}.  Including systematic uncertainties, the $\chi^2$/d.o.f. at best fit is 195/185 and 192/182 for the energy-independent and energy-dependent analysis, respectively.

\begin{table}[htp]
\caption{Summary of Systematic Uncertainties for Energy-Independent Ratio}
\label{tab:systerr_eind}
\centering
\begin{tabular}{c|c}
\hline
\hline
Condition&$\Delta R$\\
\hline
\textbf{Nuclear}\\
\hline
$\pi$ absorption&$^{+0.046}_{+0.014}$\\
\hline
$\pi$ scattering&$^{+0.059}_{-0.068}$\\
\hline
$p$ rescattering&$^{-0.076}_{+0.004}$\\
\hline
Fermi momentum&$\pm$0.012\\
\hline
\textbf{Other}\\
\hline
$M_{A}^{QE}$&$^{-0.056}_{+0.049}$\\
\hline
Bodek \& Yang&$^{+0.003}_{-0.017}$\\
\hline
$E_{\nu}$ Spectrum&$^{+0.007}_{-0.028}$\\
\hline
Crosstalk&$^{+0.042}_{-0.010}$\\
\hline
PMT 1 p.e. Resolution&$^{+0.006}_{-0.017}$\\
\hline
Birks' Constant&$^{-0.010}_{+0.037}$\\
\hline
Hit Threshold&$\pm$0.022\\
\hline
Angular Resolution&$\pm$0.011\\
\hline
MC Statistics&$\pm$0.006\\
\hline
\hline
\textbf{Total} & $^{+0.110}_{-0.126}$ \\
\hline
\hline
\end{tabular}
\end{table}

\begin{table}[htp]
\caption{Summary of Systematic Uncertainties for Energy-Dependent Ratio. The four columns $\Delta R_k$ ($k=0,1,2,3$) refer to the four neutrino energy bins defined in Section \ref{sec:analysis1}}
\label{tab:systerr_edep}
\centering
\begin{tabular}{c|cccc}
\hline
\hline
Condition&$\Delta R_{0}$&$\Delta R_{1}$&$\Delta R_{2}$&$\Delta R_{3}$\\
\hline
\hline
\textbf{Nuclear}\\
\hline
$\pi$ absorption&$^{+0.023}_{+0.068}$&$^{+0.007}_{-0.052}$&$^{+0.128}_{+0.101}$&$^{+0.197}_{-0.010}$\\
\hline
$\pi$ scattering&$^{+0.032}_{+0.013}$&$^{+0.069}_{-0.173}$&$^{+0.154}_{+0.026}$&$^{+0.013}_{-0.231}$\\
\hline
$p$ rescattering&$^{-0.071}_{+0.025}$&$^{-0.119}_{-0.019}$&$^{-0.065}_{+0.059}$&$^{-0.141}_{-0.086}$\\
\hline
Fermi momentum&$\pm$0.004&$\pm$0.021&$\pm$0.008&$\pm$0.029\\
\hline
\textbf{Other}\\
\hline
$M_{A}^{QE}$&$^{-0.038}_{+0.017}$&$^{-0.053}_{+0.048}$&$^{-0.032}_{+0.021}$&$^{-0.276}_{+0.271}$\\
\hline
Bodek \& Yang&$^{+0.007}_{-0.013}$&$^{+0.006}_{-0.021}$&$^{+0.020}_{-0.032}$&$^{-0.053}_{-0.044}$\\
\hline
$E_{\nu}$ Spectrum&$^{+0.083}_{-0.078}$&$^{+0.060}_{-0.080}$&$^{+0.188}_{-0.164}$&$^{+0.040}_{-0.221}$\\
\hline
Crosstalk&$^{+0.024}_{+0.087}$&$^{+0.031}_{-0.079}$&$^{+0.103}_{+0.075}$&$^{+0.052}_{-0.216}$\\
\hline
PMT 1 p.e. Resolution&$^{-0.005}_{-0.017}$&$^{+0.011}_{-0.025}$&$^{+0.025}_{-0.003}$&$^{-0.018}_{-0.083}$\\
\hline
Birks' Constant&$^{+0.010}_{+0.044}$&$^{-0.024}_{+0.047}$&$^{-0.005}_{+0.099}$&$^{-0.054}_{-0.135}$\\
\hline
Hit Threshold&$\pm$0.014&$\pm$0.045&$\pm$0.012&$\pm$0.168\\
\hline
Angular Resolution&$\pm$0.013&$\pm$0.001&$\pm$0.022&$\pm$0.039\\
\hline
MC Statistics&$\pm$0.006&$\pm$0.015&$\pm$0.017&$\pm$0.037\\
\hline
\hline
\textbf{Total}& $^{+0.153}_{-0.116}$ & $^{+0.130}_{-0.258}$ & $^{+0.319}_{-0.185}$ & $^{+0.386}_{-0.552}$ \\ 
\hline
\hline
\end{tabular}
\end{table}

\section{Results}
\label{sec:results}
% Editors: Ana, Lisa

Table \ref{tab:inclusive_analysis2} shows the results for the the cross section ratio $\sigma^{CC1\pi^{+}}/\sigma^{CCQE}$ for both the overall measurement and the energy-dependent measurement.  The uncertainty in the measurement due to fitting errors is labeled ``fit''; the systematic uncertainty due to the nuclear effects described in Section \ref{subsec:systematics_nucleareffects} is labeled ``nucl,'' and the uncertainty due to all other systematic effects is labeled ``syst.''  The uncertainty due to fitting errors (about 12\% for the energy-independent measurement) includes not just the statistical uncertainty, but also the error associated with degeneracies in the $p_{\mu}$ vs $\theta_{\mu}$ distributions for CC1$\pi^{+}$, CCQE, and other nonQE interactions in the various subsamples, as well as correlations among the fitting parameters ($R^{CC1\pi^{+}}$, $R^{OtherNonQE}$, and $P_{sc}$ in Table III).

\begin{table}[htp]
\caption{Cross Section Ratio}
\label{tab:inclusive_analysis2}
\centering
\begin{tabular}{c|c}
\hline
\hline
\textbf{Energy Range}&\textbf{Cross Section Ratio}\\
\textbf{(GeV)}&$R_{e} = \dfrac{\sigma^{CC1\pi^{+}}_{e}}{\sigma^{CCQE}_{e}}$\\
\hline
\hline
&\\
$>$0.00&0.734$\pm$0.086(fit)$^{+0.076}_{-0.103}$(nucl)$^{+0.079}_{-0.073}$(syst)\\
\hline
\hline
&\\
0.00-1.35&0.402$\pm$0.111(fit)$^{+0.079}_{-0.071}$(nucl)$^{+0.131}_{-0.092}$(syst)\\
1.35-1.72&1.022$\pm$0.167(fit)$^{+0.072}_{-0.217}$(nucl)$^{+0.107}_{-0.139}$(syst)\\
1.72-2.22&1.007$\pm$0.214(fit)$^{+0.209}_{-0.065}$(nucl)$^{+0.241}_{-0.173}$(syst)\\
$>$2.22&1.450$\pm$0.324(fit)$^{+0.200}_{-0.272}$(nucl)$^{+0.330}_{-0.480}$(syst)\\
\hline
\hline
\end{tabular}
\end{table}

\subsection{Comparison with Neutrino Interaction Simulation}
\label{subsec:results_comparemc}
% Editors: Ana, Lisa

\begin{figure}[ht]
\begin{center}
\includegraphics[width=0.49\textwidth]{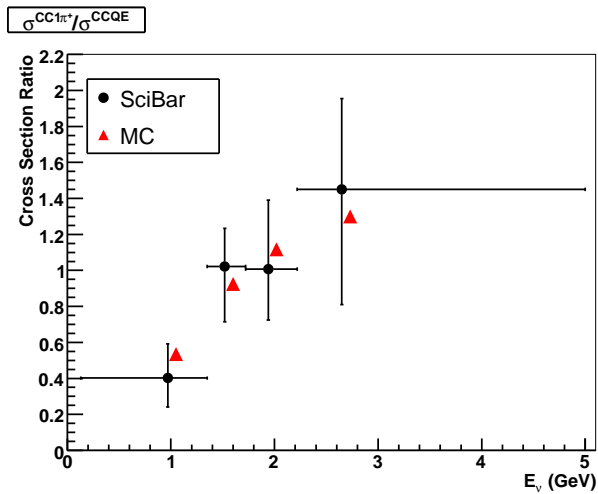}
\caption{(Color online) Comparison between the energy-dependent cross section ratio measurement and MC prediction. The vertical bars indicate the combination of the fit, nuclear, and other systematic uncertainties of the measurement. }
\label{comp-qe-mc-in}
\end{center}
\end{figure}

The MC prediction for the total cross section ratio is 0.740$\pm$0.002(stat). Figure \ref{comp-qe-mc-in} shows the comparison of the energy-dependent result with the MC prediction. The vertical bars indicate the total measurement uncertainty, and the horizontal bars indicate the neutrino energy bin width.  The results are consistent with the MC prediction.

\subsection{Comparison with Existing Results}
\label{subsec:results_comparepreviousresults}
% Editors: Ana, Lisa

In order to make the comparison with previous experimental results meaningful, the measurement is corrected to obtain the cross section ratios for an isoscalar target. SciBar is made of polystyrene ($C_8H_8$) which has 56 protons and 48 neutrons. Therefore, the scaling factor is $f = (6/7)S_p + S_n$ where $S_p$ ($S_n$) is the cross section of the $p\pi^+$ channel ($n\pi^+$) relative to the total CC1$\pi^{+}$ cross section calculated from MC. The obtained value of $f$ is 0.89.

\begin{figure}[ht]
\begin{center}
\includegraphics[width=0.49\textwidth]{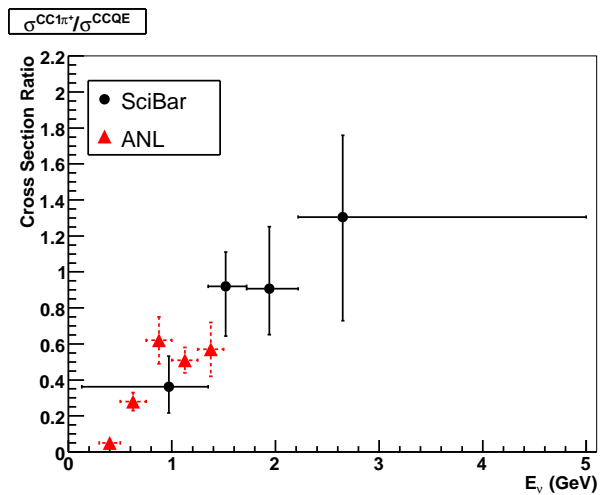}
\caption{(Color online) Energy-dependent cross section ratio measurement compared to a previous result obtained by the ANL 12-foot bubble chamber experiment. ANL CC1$\pi^{+}$ (with no hadronic invariant mass cut) and CCQE cross sections used in the comparison are taken from ~\cite{Radecky:1981fn} and \cite{Barish:1977qk}, respectively. The vertical bars indicate the combination of the fit, nuclear, and other systematic uncertainties of the measurement. }
\label{comp-qe-pre-in}
\end{center}
\end{figure}

Figure \ref{comp-qe-pre-in} shows the comparison of the energy-dependent measurement with a previous experimental result after making the correction described above.  Again, the vertical bars indicate the total uncertainty, and the horizontal bars indicate the neutrino energy bin width.  The results presented in this paper are consistent with the previous measurements.

\section{Conclusions}
\label{sec:conclusions}
% Editors: Lisa

In summary, we have studied the cross section for single charged pion production in charged-current neutrino interactions on a C$_8$H$_8$ nuclear target.  The data are collected by the SciBar detector as part of the K2K experiment, corresponding to neutrino interactions in the $\simeq 0.4-3$ GeV neutrino energy range (see Figure \ref{spectrum}). The cross section for single charged pion production in the resonance region is measured relative to the charged-current quasi-elastic cross section to avoid the large uncertainties in measuring the absolute neutrino flux.  We measure both the total cross section ratio and the cross section ratio as a function of neutrino energy.  The results are consistent with our MC prediction based on the Rein and Sehgal model and with the previous experimental result from ANL~\cite{Barish:1978pj,Radecky:1981fn,Barish:1977qk}.  
 
Compared to existing published results, and with approximately 3,000 single charged pion production interactions, this result is based on the largest event sample in this neutrino energy range to date, and the first one that uses a mostly carbon-based target. Compared to previous K2K neutrino charged-current interaction studies, this measurement provides a more detailed understanding of the interaction rates of the contributing inelastic channels, and of their energy dependence. This measurement is therefore an important contribution to the knowledge of the single pion production cross section in the few-GeV region, which is the relevant energy region for several present and future neutrino oscillation experiments.

\section{Acknowledgments}
\label{sec:acknowledgments}
% Editors: Michel, Federico
We thank the KEK and ICRR directorates for their strong support
and encouragement. K2K was made possible by the inventiveness and the
diligent efforts of the KEK-PS machine group and beam channel group.
We gratefully acknowledge the cooperation of the Kamioka Mining and Smelting Company.
This work has been supported by the Ministry of
Education, Culture, Sports, Science and Technology of the Government of Japan,
the Japan Society for Promotion of Science, the U.S. Department of Energy,
the Korea Research Foundation, the Korea Science and Engineering Foundation,
NSERC Canada and Canada Foundation for Innovation,
the Istituto Nazionale di Fisica Nucleare (Italy), the Ministerio de Educaci\'on y Ciencia and Generalitat Valenciana (Spain), the Commissariat \`a l'Energie Atomique (France), and Polish KBN grants: 1P03B08227 and 1P03B03826.


\begin{thebibliography}{999}
\bibitem{Block:1964gj}
  M.~M.~Block {\it et al.},
  %``Neutrino interactions in the CERN heavy liquid bubble chamber,''
  Phys.\ Lett.\  {\bf 12}, 281 (1964).
  %%CITATION = PHLTA,12,281;%%
%
\bibitem{Budagov:1969pw}
  I.~Budagov {\it et al.},
  %``Single pion production by neutrinos on free protons,''
  Phys.\ Lett.\  B {\bf 29}, 524 (1969).
  %%CITATION = PHLTA,B29,524;%%
%
\bibitem{Campbell:1973wg}
  J.~Campbell {\it et al.},
  %``Study of the reaction nu p $\to$ mu- pi+ p,''
  Phys.\ Rev.\ Lett.\  {\bf 30}, 335 (1973).
  %%CITATION = PRLTA,30,335;%%
%
\bibitem{Barish:1978pj}
  S.~J.~Barish {\it et al.},
  %``Study Of Neutrino Interactions In Hydrogen And Deuterium: Inelastic Charged
  %Current Reactions,''
  Phys.\ Rev.\  D {\bf 19}, 2521 (1979).
  %%CITATION = PHRVA,D19,2521;%%
%
\bibitem{Bell:1978qu}
  J.~Bell {\it et al.},
  %``Cross-Section Measurements For The Reactions Neutrino P $\to$ Mu- Pi+ P And
  %Neutrino P $\to$ Mu- K+ P At High-Energies,''
  Phys.\ Rev.\ Lett.\  {\bf 41}, 1008 (1978).
  %%CITATION = PRLTA,41,1008;%%
%
\bibitem{Bell:1978rb}
  J.~Bell {\it et al.},
  %``A Study Of The Reaction Neutrino P $\to$ Mu- Delta++ At High-Energies And
  %Comparisons With Theory,''
  Phys.\ Rev.\ Lett.\  {\bf 41}, 1012 (1978).
  %%CITATION = PRLTA,41,1012;%%
%
\bibitem{Lerche:1978cp}
  W.~Lerche {\it et al.},
  %``Experimental Study Of The Reaction Neutrino P $\to$ Mu- P Pi+. (Gargamelle
  %Neutrino Propane Experiment),''
  Phys.\ Lett.\  B {\bf 78}, 510 (1978).
  %%CITATION = PHLTA,B78,510;%%
%
\bibitem{Rein:1980wg}
  D.~Rein and L.~M.~Sehgal,
  %``Neutrino Excitation Of Baryon Resonances And Single Pion Production,''
  Annals Phys.\  {\bf 133}, 79 (1981).
  %%CITATION = APNYA,133,79;%%

%
\bibitem{Feynman:1971wr}
  R.~P.~Feynman, M.~Kislinger and F.~Ravndal,
  %``Current Matrix Elements From A Relativistic Quark Model,''
  Phys.\ Rev.\  D {\bf 3}, 2706 (1971).
  %%CITATION = PHRVA,D3,2706;%%

%
\bibitem{Schreiner:1973mj}
  P.~A.~Schreiner and F.~Von Hippel,
  %``Neutrino production of the Delta (1236),''
  Nucl.\ Phys.\  B {\bf 58}, 333 (1973).
  %%CITATION = NUPHA,B58,333;%%

%
\bibitem{Allen:1980ti}
  P.~Allen {\it et al.}  [Aachen-Bonn-CERN-Munich-Oxford Collaboration],
  %``Single Pi+ Production In Charged Current Neutrino - Hydrogen
  %Interactions,''
  Nucl.\ Phys.\  B {\bf 176}, 269 (1980).
  %%CITATION = NUPHA,B176,269;%%
%
\bibitem{Radecky:1981fn}
  G.~M.~Radecky {\it et al.},
  %``Study Of Single Pion Production By Weak Charged Currents In Low-Energy
  %Neutrino D Interactions,''
  Phys.\ Rev.\  D {\bf 25}, 1161 (1982)
  [Erratum-ibid.\  D {\bf 26}, 3297 (1982)].
  %%CITATION = PHRVA,D25,1161;%%
%
\bibitem{Kitagaki:1986ct}
  T.~Kitagaki {\it et al.},
  %``CHARGED CURRENT EXCLUSIVE PION PRODUCTION IN NEUTRINO DEUTERIUM
  %INTERACTIONS,''
  Phys.\ Rev.\  D {\bf 34}, 2554 (1986).
  %%CITATION = PHRVA,D34,2554;%%
%
\bibitem{Allen:1985ti}
  P.~Allen {\it et al.}  [Aachen-Birmingham-Bonn-CERN-London-Munich-Oxford
                  Collaboration],
  %``A Study Of Single Meson Production In Neutrino And Anti-Neutrinos Charged
  %Current Interactions On Protons,''
  Nucl.\ Phys.\  B {\bf 264}, 221 (1986).
  %%CITATION = NUPHA,B264,221;%%
%
\bibitem{Ammosov:1988xb}
  V.~V.~Ammosov {\it et al.},
  %``Study of the reaction nu p --> mu- Delta++ at energies  3-GeV to 30-GeV,''
  Sov.\ J.\ Nucl.\ Phys.\  {\bf 50}, 67 (1989)
  [Yad.\ Fiz.\  {\bf 50}, 106 (1989)].
  %%CITATION = YAFIA,50,106;%%
%
\bibitem{Grabosch:1988gw}
  H.~J.~Grabosch {\it et al.}  [SKAT Collaboration],
  %``CROSS-SECTION MEASUREMENTS OF SINGLE PION PRODUCTION IN CHARGED CURRENT
  %NEUTRINO AND ANTI-NEUTRINO INTERACTIONS,''
  Z.\ Phys.\  C {\bf 41}, 527 (1989).
  %%CITATION = ZEPYA,C41,527;%%
%
\bibitem{Allasia:1990uy}
  D.~Allasia {\it et al.},
  %``Investigation of exclusive channels in neutrino / anti-neutrino deuteron
  %charged current interactions,''
  Nucl.\ Phys.\  B {\bf 343}, 285 (1990).
  %%CITATION = NUPHA,B343,285;%%
%
\bibitem{Bandyopadhyay:2007kx}
  A.~Bandyopadhyay {\it et al.}  [ISS Physics Working Group],
  %``Physics at a future Neutrino Factory and super-beam facility,''
  arXiv:0710.4947 [hep-ph].
  %%CITATION = ARXIV:0710.4947;%%
%
\bibitem{Ahn:2006zz}
  M.~H.~Ahn {\it et al.}  [K2K Collaboration],
  %``Measurement of neutrino oscillation by the K2K experiment,''
  Phys.\ Rev.\  D {\bf 74}, 072003 (2006)
  [arXiv:hep-ex/0606032].
  %%CITATION = PHRVA,D74,072003;%%

%\cite{Aliu:2004sq}
\bibitem{Aliu:2004sq}
  E.~Aliu {\it et al.}  [K2K Collaboration],
  %``Evidence for muon neutrino oscillation in an accelerator-based
  %experiment,''
  Phys.\ Rev.\ Lett.\  {\bf 94}, 081802 (2005)
  [arXiv:hep-ex/0411038].
  %%CITATION = PRLTA,94,081802;%%

%\cite{Ahn:2002up}
\bibitem{Ahn:2002up}
  M.~H.~Ahn {\it et al.}  [K2K Collaboration],
  %``Indications of neutrino oscillation in a 250-km long-baseline experiment,''
  Phys.\ Rev.\ Lett.\  {\bf 90}, 041801 (2003)
  [arXiv:hep-ex/0212007].
  %%CITATION = PRLTA,90,041801;%%

%\cite{Ahn:2001cq}
\bibitem{Ahn:2001cq}
  S.~H.~Ahn {\it et al.}  [K2K Collaboration],
  %``Detection of accelerator produced neutrinos at a distance of 250-km,''
  Phys.\ Lett.\  B {\bf 511}, 178 (2001)
  [arXiv:hep-ex/0103001].
  %%CITATION = PHLTA,B511,178;%%

%\cite{Fukuda:2002uc}
\bibitem{Fukuda:2002uc}
  Y.~Fukuda {\it et al.},
  %``The Super-Kamiokande detector,''
  Nucl.\ Instrum.\ Meth.\  A {\bf 501}, 418 (2003).
  %%CITATION = NUIMA,A501,418;%%

%\cite{Nakayama:2004dp}
\bibitem{Nakayama:2004dp}
  S.~Nakayama {\it et al.}  [K2K Collaboration],
  %``Measurement of single pi0 production in neutral current neutrino
  %interactions with water by a 1.3-GeV wide band muon neutrino beam,''
  Phys.\ Lett.\  B {\bf 619}, 255 (2005)
  [arXiv:hep-ex/0408134].
  %%CITATION = PHLTA,B619,255;%%

%\cite{Suzuki:2000nj}
\bibitem{Suzuki:2000nj}
  A.~Suzuki {\it et al.}  [K2K Collaboration],
  %``Design, construction, and operation of SciFi tracking detector for K2K
  %experiment,''
  Nucl.\ Instrum.\ Meth.\  A {\bf 453}, 165 (2000)
  [arXiv:hep-ex/0004024].
  %%CITATION = NUIMA,A453,165;%%

\bibitem{Nitta:2004nt}
  K.~Nitta {\it et al.},
  %``The K2K SciBar detector,''
  Nucl.\ Instrum.\ Meth.\  A {\bf 535} 147 (2004).
  %[arXiv:hep-ex/0406023].
  %%CITATION = NUIMA,A535,147;%%

\bibitem{Yamamoto:2005cy}
  S.~Yamamoto {\it et al.},
  %``Design, construction, and initial performance of SciBar detector in K2K
  %experiment,''
  IEEE Trans.\ Nucl.\ Sci.\  {\bf 52} 2992 (2005).
  %%CITATION = IETNA,52,2992;%%

% \bibitem{Pla-Dalmau:2001en}
%   A.~Pla-Dalmau  [MINOS Scintillator Group],
%   %``Extruded plastic scintillator for the MINOS calorimeters,''
%   Frascati Phys.\ Ser.\  {\bf 21} 513 (2001).
%   %%CITATION = 00309,21,513;%%

\bibitem{Yoshida:2004mh}
  M.~Yoshida {\it et al.},
  %``Development of the readout system for the K2K SciBar detector,''
  IEEE Trans.\ Nucl.\ Sci.\  {\bf 51}, 3043 (2004).
  %%CITATION = IETNA,51,3043;%%

\bibitem{Hasegawa-thesis}
  M.~Hasegawa,
  Ph.D. thesis,
  Kyoto University, 2006.

\bibitem{Glazov:1993ur}
   A.~Glazov, I.~Kisel, E.~Konotopskaya, and G.~Ososkov",
   %``Filtering tracks in discrete detectors using a cellular automaton'',
   Nucl.\ Instrum.\  Meth.\ A {\bf 329} 262 (1993).
   %%CITATION = NUIMA,A329,262;%%"

\bibitem{Buontempo:1994yp}
  S.~Buontempo {\it et al.},
  %``Construction and test of calorimeter modules for the CHORUS experiment,''
  Nucl.\ Instrum.\ Meth.\  A {\bf 349}, 70 (1994).
  %%CITATION = NUIMA,A349,70;%%

%\cite{Ishii:2001sj}
\bibitem{Ishii:2001sj}
  T.~Ishii {\it et al.}  [K2K MRD GROUP Collaboration],
  %``Near muon range detector for the K2K experiment: Construction and
  %performance,''
  Nucl.\ Instrum.\ Meth.\  A {\bf 482}, 244 (2002)
  [Erratum-ibid.\  A {\bf 488}, 673 (2002)]
  [arXiv:hep-ex/0107041].
  %%CITATION = NUIMA,A482,244;%%

%\cite{Hayato:2002sd}
\bibitem{Hayato:2002sd}
  Y.~Hayato,
  %``Neut,''
  Nucl.\ Phys.\ Proc.\ Suppl.\  {\bf 112}, 171 (2002).
  %%CITATION = NUPHZ,112,171;%%

\bibitem{Rein:1987cb}
  D.~Rein,
  %``Angular Distribution In Neutrino Induced Single Pion Production
  %Processes,''
  Z.\ Phys.\  C {\bf 35}, 43 (1987).
  %%CITATION = ZEPYA,C35,43;%%

\bibitem{Smith:1972xh}
  R.~A.~Smith and E.~J.~Moniz,
  %``Neutrino Reactions On Nuclear Targets,''
  Nucl.\ Phys.\  B {\bf 43}, 605 (1972)
  [Erratum-ibid.\  B {\bf 101}, 547 (1975)].
  %%CITATION = NUPHA,B43,605;%%

\bibitem{Gluck:1994uf}
  M.~Gluck, E.~Reya and A.~Vogt,
  %``Dynamical Parton Distributions Of The Proton And Small X Physics,''
  Z.\ Phys.\  C {\bf 67}, 433 (1995).
  %%CITATION = ZEPYA,C67,433;%%

\bibitem{Bodek:2002vp}
  A.~Bodek and U.K.~Yang,
  %"Modeling deep inelastic cross sections in the few GeV region",
  Nucl. Phys. Proc. Suppl. {\bf 112} 70 (2002).
  %%CITATION = HEP-EX 0203009;%%"

%\cite{Sjostrand:1993yb}
\bibitem{Sjostrand:1993yb}
  T.~Sjostrand,
  %``High-energy physics event generation with PYTHIA 5.7 and JETSET 7.4,''
  Comput.\ Phys.\ Commun.\  {\bf 82}, 74 (1994).
  %%CITATION = CPHCB,82,74;%%

%\cite{Nakahata:1986zp}
\bibitem{Nakahata:1986zp}
  M.~Nakahata {\it et al.}  [KAMIOKANDE Collaboration],
  %``ATMOSPHERIC NEUTRINO BACKGROUND AND PION NUCLEAR EFFECT FOR KAMIOKA NUCLEON
  %DECAY EXPERIMENT,''
  J.\ Phys.\ Soc.\ Jap.\  {\bf 55}, 3786 (1986).
  %%CITATION = JUPSA,55,3786;%%

\bibitem{Rein:1983pf}
  D.~Rein and L.~M.~Sehgal,
  %``Coherent Pi0 Production In Neutrino Reactions,''
  Nucl.\ Phys.\  B {\bf 223}, 29 (1983).
  %%CITATION = NUPHA,B223,29;%%

\bibitem{Hasegawa:2005td}
  M.~Hasegawa {\it et al.}  [K2K Collaboration],
  %``Search for coherent charged pion production in neutrino carbon
  %interactions,''
  Phys.\ Rev.\ Lett.\  {\bf 95}, 252301 (2005)
  [arXiv:hep-ex/0506008].
  %%CITATION = PRLTA,95,252301;%%

%\cite{Brun:1987ma}
\bibitem{Brun:1987ma}
  R.~Brun, F.~Bruyant, M.~Maire, A.~C.~McPherson and P.~Zanarini,
  %``GEANT3,''
  CERN Report CERN-DD/EE/84-1 (1987).
  %%CITATION = CERN-DD/EE/84-1;%%

%\cite{Zeitnitz:1994bs}
\bibitem{Zeitnitz:1994bs}
  C.~Zeitnitz and T.~A.~Gabriel,
  %``The GEANT - CALOR interface and benchmark calculations of ZEUS test
  %calorimeters,''
  Nucl.\ Instrum.\ Meth.\  A {\bf 349}, 106 (1994).
  %%CITATION = NUIMA,A349,106;%%

\bibitem{birks}
   J.~Birks,
   \textit{Theory and Practice of Scintillation Counting},
   Pergamon Press, 1964.

\bibitem{Yao:2006px}
  W.~M.~Yao {\it et al.}  [Particle Data Group],
  %``Review of particle physics,''
  J.\ Phys.\ G {\bf 33}, 1 (2006).
  %%CITATION = JPHGB,G33,1;%%

\bibitem{Barish:1977qk}
  S.~J.~Barish {\it et al.},
  %``Study Of Neutrino Interactions In Hydrogen And Deuterium. 1. Description Of
  %The Experiment And Study Of The Reaction Neutrino D $\to$ Mu- P P(S),''
  Phys.\ Rev.\  D {\bf 16}, 3103 (1977).
  %%CITATION = PHRVA,D16,3103;%%

\bibitem{Bonetti:1977cs}
  S.~Bonetti {\it et al.},
  %``Study Of Quasielastic Reactions Of Neutrino And Anti-Neutrino In
  %Gargamelle,''
  Nuovo Cim.\  A {\bf 38}, 260 (1977).
  %%CITATION = NUCIA,38A,260;%%

\bibitem{ggm}
  S.~Ciampolillo {\it et al.}  [Gargamelle Neutrino Propane Collaboration],
  %``Total Cross-Section For Neutrino Charged Current Interactions At 3-Gev And
  %9-Gev,''
  Phys.\ Lett.\  B {\bf 84}, 281 (1979);
  %%CITATION = PHLTA,B84,281;%%
  N.~Armenise {\it et al.},
  %``Charged Current Elastic Anti-Neutrino Interactions In Propane,''
  Nucl.\ Phys.\  B {\bf 152}, 365 (1979).
  %%CITATION = NUPHA,B152,365;%%

\bibitem{Belikov:1983kg}
  S.~V.~Belikov {\it et al.},
  %``Quasielastic Neutrino And Anti-Neutrinos Scattering: Total Cross-Sections,
  %Axial Vector Form-Factor,''
  Z.\ Phys.\  A {\bf 320}, 625 (1985).
  %%CITATION = ZEPYA,A320,625;%%

%\cite{Gran:2006jn}
\bibitem{Gran:2006jn}
  R.~Gran {\it et al.}  [K2K Collaboration],
  %``Measurement of the quasi-elastic axial vector mass in neutrino oxygen
  %interactions,''
  Phys.\ Rev.\  D {\bf 74}, 052002 (2006)
  [arXiv:hep-ex/0603034].
  %%CITATION = PHRVA,D74,052002;%%

\bibitem{Ingram:1982bn}
  C.~H.~Q.~Ingram {\it et al.},
  %``Measurement Of Quasielastic Scattering Of Pions From O-16 At Energies
  %Around The Delta (1232) Resonance,''
  Phys.\ Rev.\  C {\bf 27}, 1578 (1983).
  %%CITATION = PHRVA,C27,1578;%% 
%
\bibitem{Jeon-thesis}
  E.~J.~Jeon,
  Ph.D. thesis, The Graduate University for Advanced Studies (Sokendai), 2003.

%\cite{Moniz:1971mt}
\bibitem{Moniz:1971mt}
  E.~J.~Moniz, I.~Sick, R.~R.~Whitney, J.~R.~Ficenec, R.~D.~Kephart and W.~P.~Trower,
  %``Nuclear fermi momenta from quasielastic electron scattering,''
  Phys.\ Rev.\ Lett.\  {\bf 26}, 445 (1971).
  %%CITATION = PRLTA,26,445;%%







% \bibitem{gabriel}
%   G.~Jover Manas,
%   %``Measurement of the cross section of the multipion interaction in a carbon target using the SciBar detector at K2K experiment,''
%   Ph.D. thesis (in preparation),
%   Barcelona University.



% %\cite{Kitagaki:1990vs}
% \bibitem{Kitagaki:1990vs}
%   T.~Kitagaki {\it et al.},
%   %``Study Of Neutrino D $\to$ Mu- P P(S) And Neutrino D $\to$ Mu- Delta++
%   %(1232) N(S) Using The Bnl 7-Foot Deuterium Filled Bubble Chamber,''
%   Phys.\ Rev.\  D {\bf 42}, 1331 (1990).
%   %%CITATION = PHRVA,D42,1331;%%
% 
% %\cite{Pohl:1979zm}
% \bibitem{Pohl:1979zm}
%   M.~Pohl {\it et al.}  [GARGAMELLE NEUTRINO PROPANE COLLABORATION
%                   Collaboration],
%   %``Experimental Study Of The Reaction Neutrino N $\to$ Mu- P,''
%   Lett.\ Nuovo Cim.\  {\bf 26}, 332 (1979).
%   %%CITATION = NCLTA,26,332;%%


%

%

% \bibitem{Kitagaki:1990vs}
%   T.~Kitagaki {\it et al.},
%   %``Study Of Neutrino D $\to$ Mu- P P(S) And Neutrino D $\to$ Mu- Delta++
%   %(1232) N(S) Using The Bnl 7-Foot Deuterium Filled Bubble Chamber,''
%   Phys.\ Rev.\  D {\bf 42}, 1331 (1990).
%   %%CITATION = PHRVA,D42,1331;%%
% 
% \bibitem{Pohl:1979zm}
%   M.~Pohl {\it et al.}  [GARGAMELLE NEUTRINO PROPANE COLLABORATION
%                   Collaboration],
%   %``Experimental Study Of The Reaction Neutrino N $\to$ Mu- P,''
%   Lett.\ Nuovo Cim.\  {\bf 26}, 332 (1979).
%   %%CITATION = NCLTA,26,332;%%


\end{thebibliography}
\end{document}